\documentclass[fleqn,usenatbib]{mnras}

\usepackage{newtxtext,newtxmath}

\usepackage[T1]{fontenc}
\usepackage{graphicx} 
\usepackage{graphicx}
\usepackage{subcaption} 

\DeclareRobustCommand{\VAN}[3]{#2}
\let\VANthebibliography\thebibliography
\def\thebibliography{\DeclareRobustCommand{\VAN}[3]{##3}\VANthebibliography}

\usepackage{graphicx}	
\usepackage{amsmath}	

\usepackage{multirow}

\usepackage{graphicx}
\usepackage{amsmath}
\usepackage{natbib}
\usepackage{xcolor}

\newcommand{\colorbold}[2]{\textcolor{#1}{\textbf{#2}}}

\title[Pseudo-nulling Pulsar PSR J1820-0509]{Single-Pulse Study of the Pseudo-nulling Pulsar PSR J1820-0509 Based on FAST Observations}

\author[Z.F. Tu et al.]{
Zefeng Tu,$^{1}$
Rushuang Zhao,$^{1}$\thanks{E-mail: 201907007@gznu.edu.cn}
Hui Liu,$^{1}$
Biping Gong,$^{2}$
D. Li,$^{6,3,4}$
P. Wang,$^{3,4,5}$
Chenchen Miao,$^{7}$
\newauthor
Q.J. Zhi,$^{11}$
S.J. Dang$^{1}$
S.D. Wang,$^{1}$
Q. Zhou,$^{1}$
Z.J. Zhang,$^{1}$
Xu Zhu,$^{8}$
R.W. Tian,$^{9}$
H.W.Xu,$^{10}$
\newauthor
Yi Zhou,$^{1}$
and D.Y. Yan$^{1}$
\\
$^{1}$Guizhou Normal University, Guiyang 550001, People’s Republic of China\\
$^{2}$Huazhong University of Science and Technology, School of Physics, Wuhan 430074, China\\
$^{3}$National Astronomical Observatories, Chinese Academy of Sciences, 20A Datun Road, Chaoyang District, Beijing 100101, People’s Republic of China\\
$^{4}$State Key Laboratory of Radio Astronomy and Technology, Chinese Academy of Sciences,\\
\hspace{0.5em}A20 Datun Road, Chaoyang District, Beijing 100101, People’s Republic of China\\
$^{5}$Institute for Frontiers in Astronomy and Astrophysics, Beijing Normal University, Beijing 102206, People’s Republic of China\\
$^{6}$Department of Astronomy, Tsinghua University, Beijing 100084, People’s Republic of China\\
$^{7}$Qilu Normal University, College of Physics and Electronic Engineering, No. 2 Wenbo Road, Zhangqiu District, Jinan, China\\
$^{8}$Guangzhou University, School of Physics and Information Engineering, Guangzhou 510006, China\\
$^{9}$Guangxi University, College of Physics and Electronic Information, Nanning 530004, China\\
$^{10}$Central China Normal University, College of Physics, Wuhan 430079, China\\
$^{11}$Guizhou University, Guiyang 550025, People’s Republic of China
}

\date{Accepted XXX. Received YYY; in original form ZZZ}

\pubyear{\the\year{}}

\begin{document}

\label{firstpage}
\pagerange{\pageref{firstpage}--\pageref{lastpage}}
\maketitle

\begin{abstract}
Using two observations from the Five-hundred-meter Aperture Spherical radio Telescope (FAST), we performed a detailed single-pulse analysis of the high-nulling pulsar PSR~J1820$-$0509. We find an exceptionally high nulling fraction of about $81.78\%$, significantly exceeding previous results from Parkes observations. The single-pulse energy distribution shows a clear bimodal structure, consistent with classical nulling behavior. However, stacking the identified null pulses reveals a statistically significant residual profile above the noise level, indicating that these nulls correspond to a very weak emission state rather than a complete cessation of radio emission.In addition, PSR~J1820$-$0509 exhibits clustered burst activities lasting several hundred rotation periods, with significant quasi-periodicities at characteristic timescales of $1191 \pm 81$ and $590 \pm 15$ pulse periods in the two observations. Based on the temporal clustering and integrated profile morphology, we identify three distinct emission modes (Modes~A, B, and C) and a pseudo-null state (Mode~D). These modes show systematic differences in pulse morphology, polarization, and energy statistics.
The pulse width--energy relations for all detectable modes display significant transitions between low- and high-energy regimes. The energy distributions of Modes~A and C are well described by lognormal functions, while Mode~B follows a composite distribution of Gaussian and lognormal components. These results suggest that the radio emission is governed by multiple quasi-stable magnetospheric states. The detection of weak emission during pseudo-nulls, together with systematic mode differences, supports the interpretation that pulsar nulling reflects transitions between different magnetospheric activity levels rather than a complete shutdown of emission.
\end{abstract}




\section{Introduction}
Since the discovery of the first pulsar by Bell in 1967, the number of pulsars discovered has continued to increase \citep{hewish1968pulsars}. The ongoing improvement in observational equipment sensitivity has greatly contributed to expanding the population of new pulsar candidates \citep{chen2023meerkat, han2021fast} and further revealing various pulsar observational phenomena \citep{yan2024dwarf}. This progress has provided opportunities for a deeper understanding of single-pulse characteristics. Numerous intriguing single-pulse phenomena have been observed in pulsar studies, such as nulling \citep{van2002null,janssen2004intermittent,redman2005pulsar} and mode changing \citep{backer1970pulsar}.

The nulling phenomenon is characterized by the cessation of entire pulse emission for a period of time, ranging from several pulse periods to several hours \citep{backer1970pulsar,rankin1986toward,wang2007pulsar,ng2020discovery}. Nulling is generally regarded a random switching process between the null state and the emission state. However, subsequent studies have revealed that some pulsars exhibit periodic nulling \citep[e.g.,][]{herfindal2007periodic}. To date, more than 200 pulsars have shown nulling phenomena \citep{wang2020two}, and \citet{basu2020periodic} summarized about 70 pulsars exhibiting ``periodic modulation,'' which includes two types: ``periodic nulling'' and ``periodic amplitude modulation.'' However, the physical mechanism behind nulling remains poorly understood.

The stability of the average profile is fundamental for pulsar timing and template matching. For most sources, integrating several hundred single pulses results in a stable average profile; however, the number of pulses required varies significantly between different pulsars \citep{helfand1975observations,rathnasree1995approach, liu2012profile}. On the other hand, mode change is another form of abrupt variation, where the average pulse profile suddenly transitions between two (or more) quasi-stable states \citep{wang2007pulsar}. The mode change phenomenon was first reported by \citet{backer1970peculiar} in the pulsar PSR~B1237+25. The mode-dependent average profiles have also been widely used to study the emission geometry of pulsars \citep{backer1970peculiar}. Most pulsars with mode change also exhibit multiple components. Many of these pulsars show subpulse drifting and nulling \citep{van2002null,janssen2004intermittent,redman2005pulsar}. Mode change can influence pulsar quasi-periodic modulation, drifting, microstructure, and polarization characteristics \citep{rathnasree1995approach,rankin1974individual,rankin2013drifting}, representing a fundamental change in the emission process and reflecting the complexity of the pulsar emission mechanism \citep{wang2007pulsar}.

PSR~J1820$-$0509 is a pulsar discovered during the second reprocessing phase of the Parkes multibeam survey \citep{lorimer2006parkes} and is known for its high nulling fraction. It has a relatively short rotation period but is older (with a characteristic age of 5.7~Myr). Most of the time, it remains in the null state, only exhibiting brief burst states (with an average burst interval of $< 1$~minute). N. Wang et al. \citep{wang2007pulsar} determined that its nulling fraction is as high as 67\%, and its power spectrum does not show any obvious periodicity, nor was any evidence of mode change found. However, A. Y. Yang et al. \citep{yang2014new} suggested that there may be some periodicity in its nulling behavior. 
Due to limitations in observational sensitivity, the fine structure of its single pulses remains unresolved. Therefore, conducting high-sensitivity observations of this pulsar to reveal its unique single-pulse behavior is of particular importance.
This paper presents the single-pulse observation of PSR~J1820$-$0509 at a central frequency of 1250~MHz using the 500-meter aperture spherical radio telescope (FAST). We used FAST to study this pulsar's nulling, mode change, and bright pulses, and explore the close relationships between these phenomena. The structure of this paper is as follows. \S2 presents the observation of PSR~J1820$-$0509 with FAST and the data processing methods. \S3 provides the analysis and results for this pulsar. The discussion and conclusions are presented in \S4.

\section{Observations and Data Analysis}

Both observations of PSR~J1820$-$0509 were conducted using the FAST, utilizing the 19-beam receiver, with a frequency range from 1.05 to 1.45~GHz. The observational data were collected and processed through the multi-channel backend of FAST, which is based on the ROACH2 signal processor. This system supports parallel observations of various scientific targets, including pulsars, HI, and fast radio bursts \citep{di2018considerations}. The time resolution for both Observation~I and Observation~II was 49.152~$\mu$s, and the frequency resolution was 0.122~MHz. The observation data were recorded in PSRFITS format with 4096 frequency channels, containing four types of polarization data (Stokes parameters). Specific details of the observations, including the nulling fraction and the number of burst states, are provided in Table~\ref{tab:obs_summary} .Due to RFI in the time domain, we selected 17082 single pulses (fifth column of the table) from the total sample as the dataset for this study.

\begin{table}
    \centering
    \caption{Single-pulse observations of PSR~J1820$-$0509 from two sessions. The total duration (in seconds), total number of pulses, number of valid pulses, nulling fraction for each observation are shown in columns 3–6, respectively.}
    \label{tab:obs_summary}
    \resizebox{0.5\textwidth}{!}{ 
    \begin{tabular}{ccccccc}
        \hline
        Observation & Date & Duration (s) & All Pulses & Effective Pulses & NF (\%) \\
        \hline
        I & 2023-05-20 & 2934 & 8706 & 8557 & 77.70  \\
        II & 2024-10-17 & 2933 & 8702 & 8525 & 85.86  \\
        \hline
    \end{tabular}
    }
\end{table}

The pulsar data were dedispersed using \textsc{dspsr} \citep{van2011dspsr} and stored as single-pulse stacks with 1024 phase bins per pulse, with a period of 0.337 s. Radio-frequency interference (RFI) was removed using standardized tools in the \textsc{psrchive} package \citep{hotan2004psrchive}. The resulting cleaned dataset comprises pulses (Figure~\ref{fig:imshow}), preserving both total intensity (Stokes I) and full polarization information (Stokes QUV) for individual pulses. This comprehensive dataset provides a robust foundation for detailed analyses of pulse morphology, energy distributions, and polarization evolution characteristics.



\section{Results and Analysis}
\subsection{Nulling}
\begin{figure}
    \centering
    \includegraphics[width=\columnwidth,height=0.6\textheight,keepaspectratio]{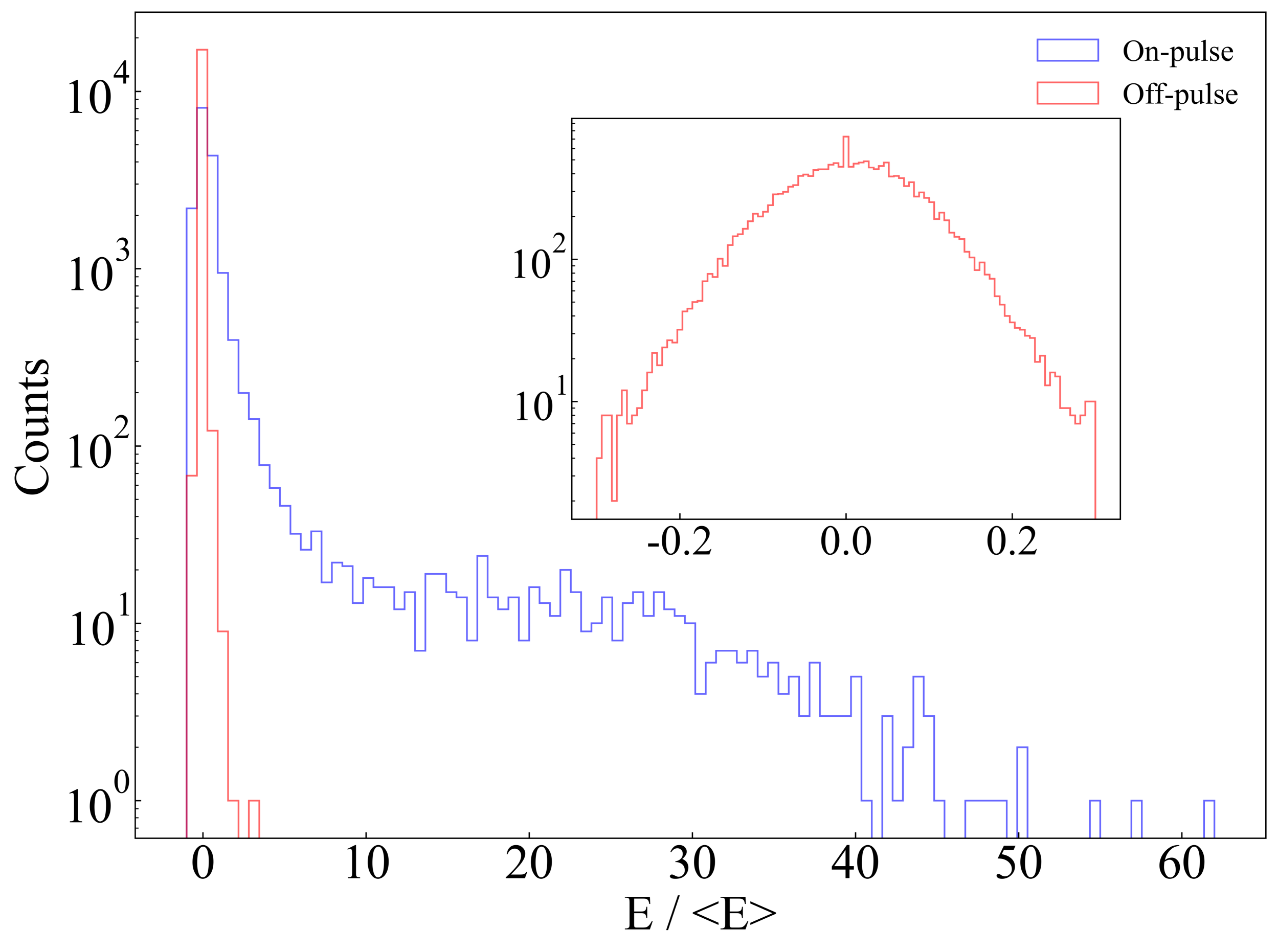}
    \caption{Single-pulse relative energy (E/<E>) distributions of PSR J1820$-$0509 in the on-pulse and off-pulse windows. The blue curve represents the on-pulse region, while the red curve corresponds to the off-pulse region, with identical phase ranges selected for both. The inset shows a zoomed view of the off-pulse energy distribution. The on-pulse distribution exhibits a bimodal structure, corresponding to null pulses and normal emission, respectively.}
    \label{fig:All_pulsar_energy}
\end{figure}

Pulsar nulling usually shows as the complete disappearance of radiation in a single pulse or a sequence of consecutive pulses. In this pulsar, the nulling fraction is about 81.78\%, which is higher than the 67\% previously measured with the Parkes telescope \citep{wang2007pulsar}.
Single$-$pulse energy analysis of our two FAST observations reveals that PSR J1820$-$0509 exhibits a clear bimodal on$-$pulse energy distribution in both datasets (Figure~\ref{fig:All_pulsar_energy}). This result aligns with the classical nulling$-$pulsar energy distribution model characterized by two peaks: one near zero energy (null pulses) and another around the mean energy (normal emission). 

In-depth analysis of the null pulses from PSR~J1820$-$0509 reveals that its ``null pulses'' are not true cessation of emission. We calculate the energy $E_{\rm p}$ in the on-pulse region of each single pulse, with its uncertainty $\sigma_{E_{\rm p, on}} = \sqrt{n_{\rm on}} \, \sigma_{\rm off}$(where $\sigma_{\rm off}$ is the root mean square of the off-pulse region noise, and $n_{\rm on}$ is the number of on-pulse phase bins)\citep{zhao2017tmrt}. 
Following the method of \citealt{bhattacharyya2010investigation} for PSR~B0818$-$41, pulses satisfying $E_{\rm p} < 3 \times \sigma_{E_{\rm p, on}}$ are defined as candidate null pulses.
We further refine our null-pulse identification method. Specifically, we first directly examine whether the peak intensity of each single pulse within the on-pulse window exceeds a threshold. To increase sensitivity to narrow, weak emission features, we compute the local mean intensity over adjacent groups of 4 phase bins within the on-pulse window, and test whether this locally averaged intensity exceeds the corresponding threshold. For both cases, the detection thresholds are set at five times the noise standard deviation, recalculated according to the number of bins included. Under these criteria, a single pulse is classified as an astrophysical emission pulse if either its peak intensity or local average intensity exceeds the threshold; otherwise, it is classified as a null pulse\citep{liu2015single, palliyaguru2023single}.

\begin{figure}
    \centering
    \includegraphics[width=\columnwidth,height=0.6\textheight,keepaspectratio]{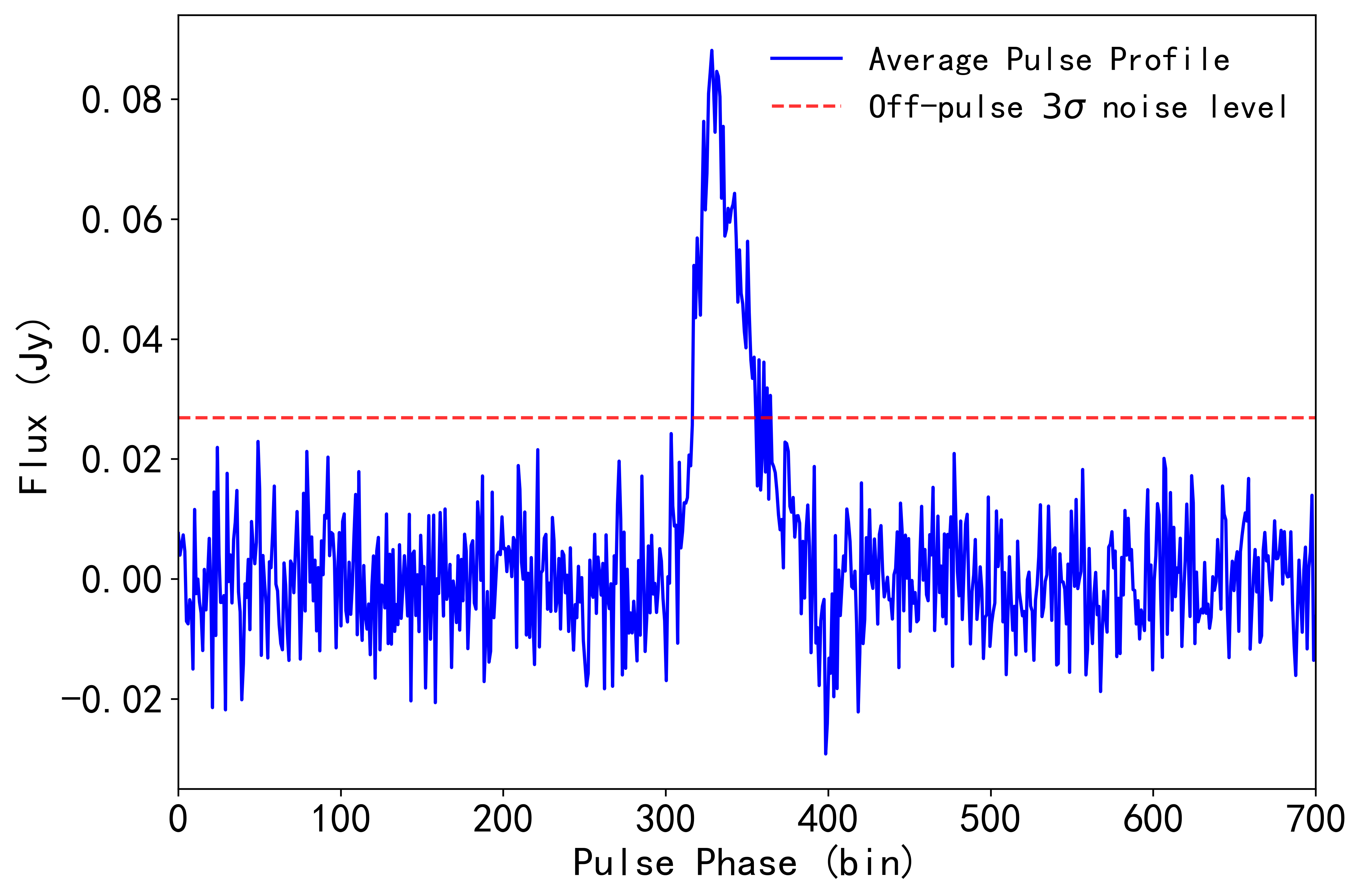}
    \caption{Average profile obtained by stacking all selected null pulses (blue curve). The red dashed line indicates the $3\sigma$ noise level of the off-pulse region. The result shows that the stacked signal is significantly above the noise, demonstrating the widespread presence of weak emission components within the nulls.}
    \label{fig:All_null}
\end{figure}

\begin{figure}
    \centering
    \includegraphics[width=\columnwidth,height=60mm]{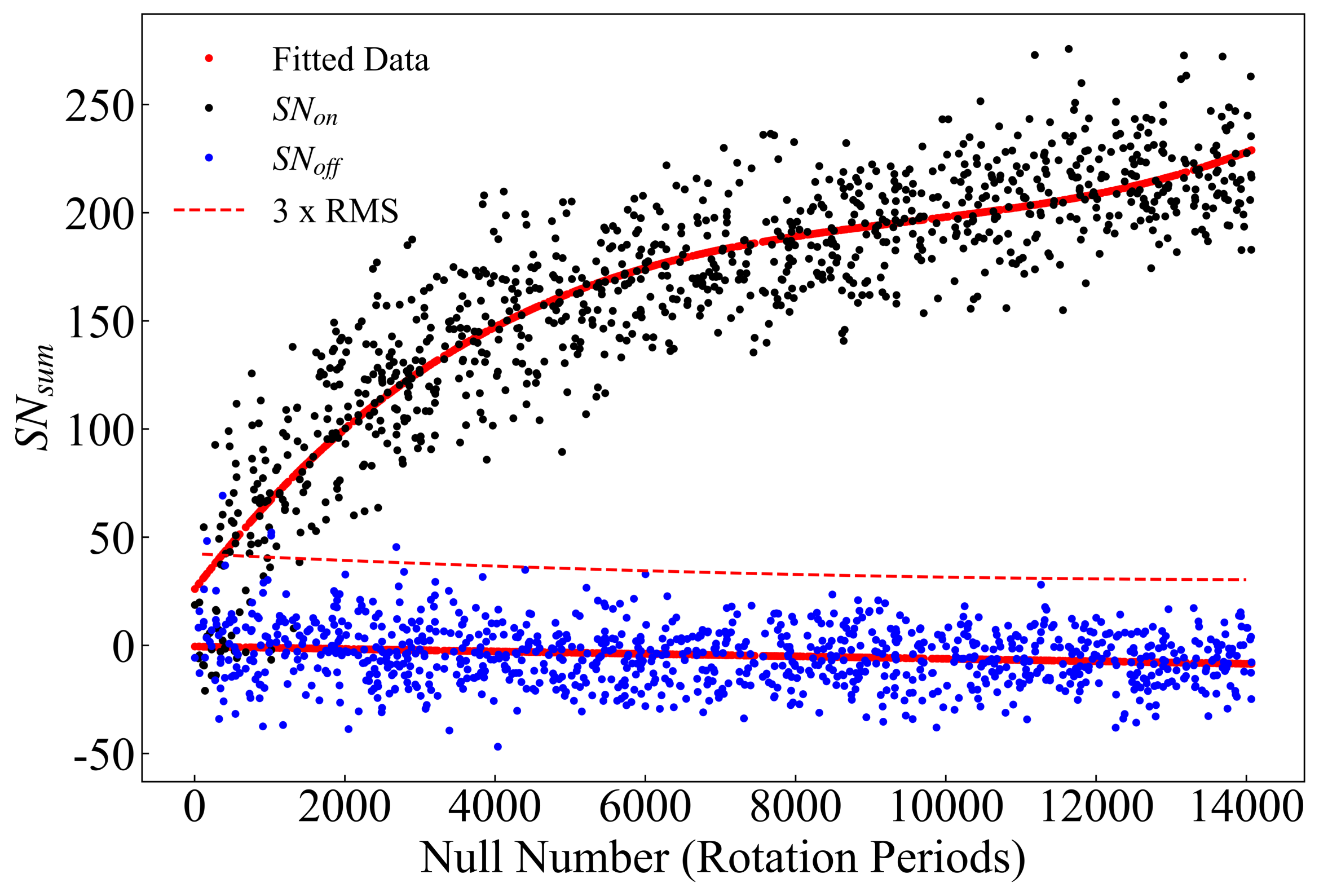}
    \caption{Cumulative signal-to-noise ratio distribution of null pulses. Black points represent the on-pulse region, and blue points represent the off-pulse region. The red dashed line represents three times the RMS of $S/N_{\rm off}$, where the RMS is derived from the fitting curve shown in Figure~\ref{fig:4SN_RMS}.The S/N{\rm sum} in the on-pulse region continues to increase with the number of stacked pulses, providing further evidence that systematic weak emission is present within the nulls.}
    \label{fig:on-pulse and off-pulse energy}
\end{figure}

\begin{figure}
    \centering
    \includegraphics[width=\columnwidth]{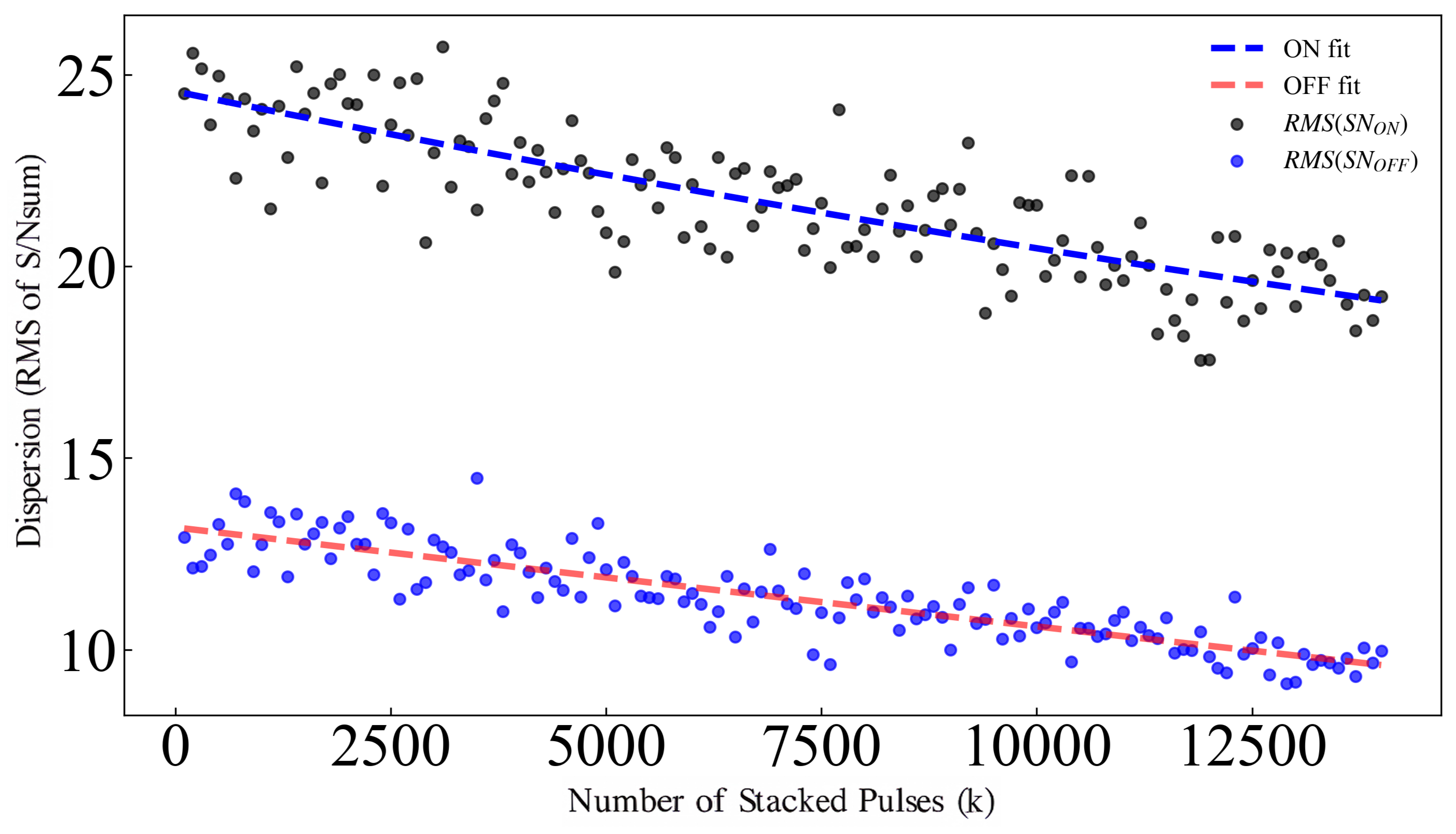}
    \caption{The figure shows the root mean square (RMS) of $\mathrm{S/N}_{\mathrm{sum}}(k)$ as a function of the number of stacked pulses for different integration scales. Each data point is derived from 200 bootstrap realizations with replacement. The black points represent the on-pulse data, while the blue points correspond to the off-pulse data. The blue dashed line indicates the fitting result for the on-pulse data, and the red dashed line represents the fitting result for the off-pulse data. Both fitted trends decrease with increasing pulse number, indicating that the stacked null-pulse profile becomes progressively more stable as the sample size increases.
}.
    \label{fig:4SN_RMS}
\end{figure}

Furthermore, we performed a stacking of all selected null pulses, and found that the integrated profile was significantly higher than three times the standard deviation of the off-pulse region (Figure~\ref{fig:All_null}). This indicates that some ``null pulses'' actually carry radiation below the detection threshold of single pulses. Similar phenomena have been reported in PSR~B0826$-$34 \citep{bhattacharyya2008results} and PSR~J1853+0505 \citep{young2015long}. 
To investigate the statistical behavior of the S/N of the integrated profile during the stacking of null pulses, We first compiled all observed null pulses into a set $N$. The on-pulse window was defined as the $3\sigma$ width of the integrated pulse profile, and an off-pulse window of the same width, located in a non-emission phase region, was selected for comparison. We then drew pulses from $N$ sequentially in a random order and constructed the integrated profile through cumulative averaging. After stacking the first $k$ pulses, we obtained the corresponding $k$-pulse averaged profile and calculated the signal-to-noise ratio of the profile, $S/N_{\mathrm{sum}}$, within both the on-pulse and off-pulse windows (See Equations~(1) and~(2) for the explicit definition). 
For each $k$, we recorded $S/N_{\mathrm{on}}(k)$ and $S/N_{\mathrm{off}}(k)$, shown as black and blue points, respectively. This procedure was repeated until all null pulses in $N$ were included, yielding the evolution of the signal-to-noise ratio of the integrated profile as a function of the number of stacked pulses, as shown in Figure~\ref{fig:on-pulse and off-pulse energy}.
The blue points represent the signal-to-noise ratio measured from the stacked profile after $k$ null pulses, evaluated within an equal-width off-pulse window located in a non-emission phase region. These points therefore represent the noise-only baseline fluctuation of the stacked-profile $S/N$.
The red dashed line represents three times the RMS of $S/N_{\rm off}$, where the RMS is derived from the fitting curve shown in Figure~\ref{fig:4SN_RMS}, serving as a noise-based significance criterion: if $S/N_{\mathrm{on}}(k)$ exceeds this line, the on-pulse window shows a statistically significant signal in the stacked profile that is unlikely to be produced by noise alone.
The black points represent the signal-to-noise ratio $S/N_{\mathrm{on}}(k)$ of the stacked profile after $k$ null pulses, measured within the on-pulse window defined by the $3\sigma$ width of the integrated profile. They quantify whether a real weak residual emission builds up in the on-pulse phase during the null state. As shown in Figure~\ref{fig:on-pulse and off-pulse energy}, a clearly identifiable pulse profile emerges after stacking $\sim 500$ null pulses.
The integrated signal-to-noise ratio (S/N{\rm sum}) of the on-pulse region (defined as the sum of all energy in the on-pulse region divided by the rms of an off-pulse region with the same number of bins) continued to increase, while the off-pulse region remained stable around zero , confirming that weak emission is commonly present in null pulses.
\begin{equation}
S/N_{\mathrm{on}} = \frac{\sum_{i \in \mathrm{on}} I_i}{\sigma_{\mathrm{off}}}
\end{equation}
\begin{equation}
S/N_{\mathrm{off}} = \frac{\sum_{i \in \mathrm{off}} I_i}{\sigma_{\mathrm{off}}}
\end{equation}
where $I_i$ is the intensity of the $i$-th phase bin of the pulse profile. 
The symbols $\sum_{i \in \mathrm{on}}$ and $\sum_{i \in \mathrm{off}}$ indicate 
the summation over the phase bins within the on-pulse and off-pulse windows, 
respectively. $\sigma_{\mathrm{off}}$ denotes the root-mean-square (rms) noise 
estimated from the off-pulse region of the profile.

Figure~\ref{fig:4SN_RMS}  shows how large the scatter in $S/N_{\mathrm{sum}}(k)$ can be at a fixed stack size $k$ when different random subsets of null pulses are used for stacking. In other words, it characterizes the stability of the random stacking result, rather than the signal strength itself.

We used sampling with replacement because it better captures the statistical scatter under repeated random resampling. For each fixed $k$, we randomly selected $k$ null pulses from the full null-pulse sample with replacement, constructed the corresponding stacked profile, and calculated $S/N_{\mathrm{sum}}(k)$. This procedure was repeated 200 times for the same $k$, yielding 200 values of $S/N_{\mathrm{sum}}(k)$. The scatter at that stack size was then quantified by the rms (equivalently, the standard deviation) of these 200 values. Repeating this procedure for increasing $k$ in steps of 100 pulses gives the points shown in Figure~\ref{fig:4SN_RMS}.
As shown in this figure, the black points correspond to the RMS  of $S/N_{\rm on}(k)$, while the blue points correspond to the RMS of $S/N_{\rm off}(k)$. The dispersion of both $S/N_{\mathrm{on}}(k)$ and $S/N_{\mathrm{off}}(k)$decreases with increasing $k$.
This behavior indicates that, as more null pulses are included in the averaging, the stacked profile becomes progressively less sensitive to the particular pulses selected in each random sample. Consequently, the random-stacking result becomes increasingly stable.

Figure~\ref{fig:4SN_RMS} therefore provides a statistical supplement to Figure~\ref{fig:on-pulse and off-pulse energy}. While Figure~\ref{fig:on-pulse and off-pulse energy} shows that the stacked signal in the on-pulse window increases with increasing $k$, Figure~\ref{fig:4SN_RMS} demonstrates that the scatter associated with the random stacking decreases as $k$ increases. The two figures are thus fully consistent: as the number of averaged null pulses increases, the signal growth seen in Figure~\ref{fig:on-pulse and off-pulse energy} becomes progressively more reliable and is unlikely to be caused by a small number of exceptional pulses selected by chance. Instead, it supports the presence of a weak but persistent emission component during the null state, which becomes detectable only after averaging a sufficiently large number of null pulses. This result is consistent with the study of the high-null-fraction pulsar PSR~B0818$-$41 \citep{bhattacharyya2010investigation}, and jointly challenges the traditional physical definition of nulling.

\subsection{Energy Statistical Analysis}

\begin{figure*} 
    \centering
    \begin{subfigure}{0.30\textwidth}
        \includegraphics[width=\columnwidth,height=0.6\textheight,keepaspectratio]{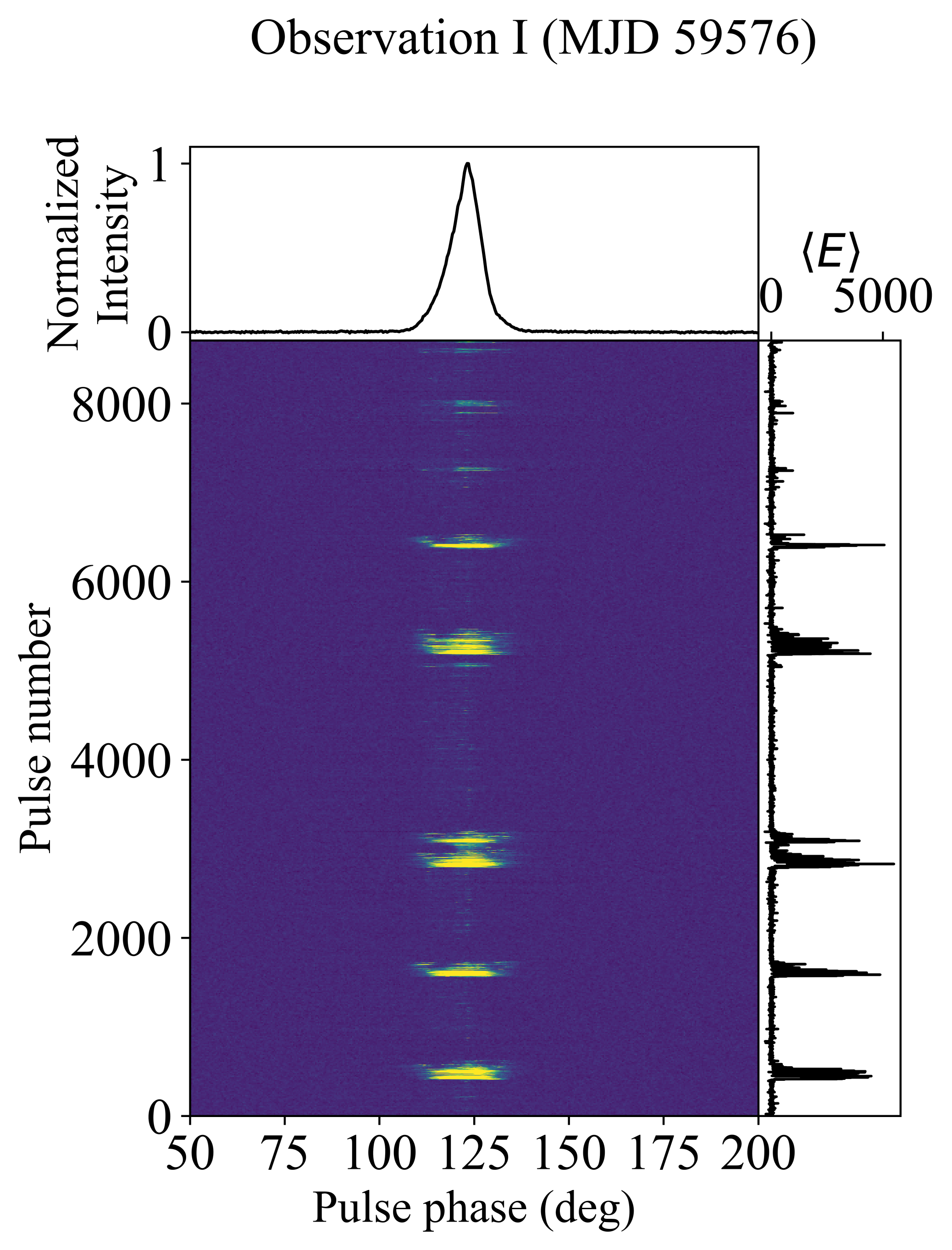}
        \caption{}
        \label{fig:imshow I}
    \end{subfigure}
    \hfill
    \begin{subfigure}{0.30\textwidth}
    \centering
        \includegraphics[width=\columnwidth,height=0.6\textheight,keepaspectratio]{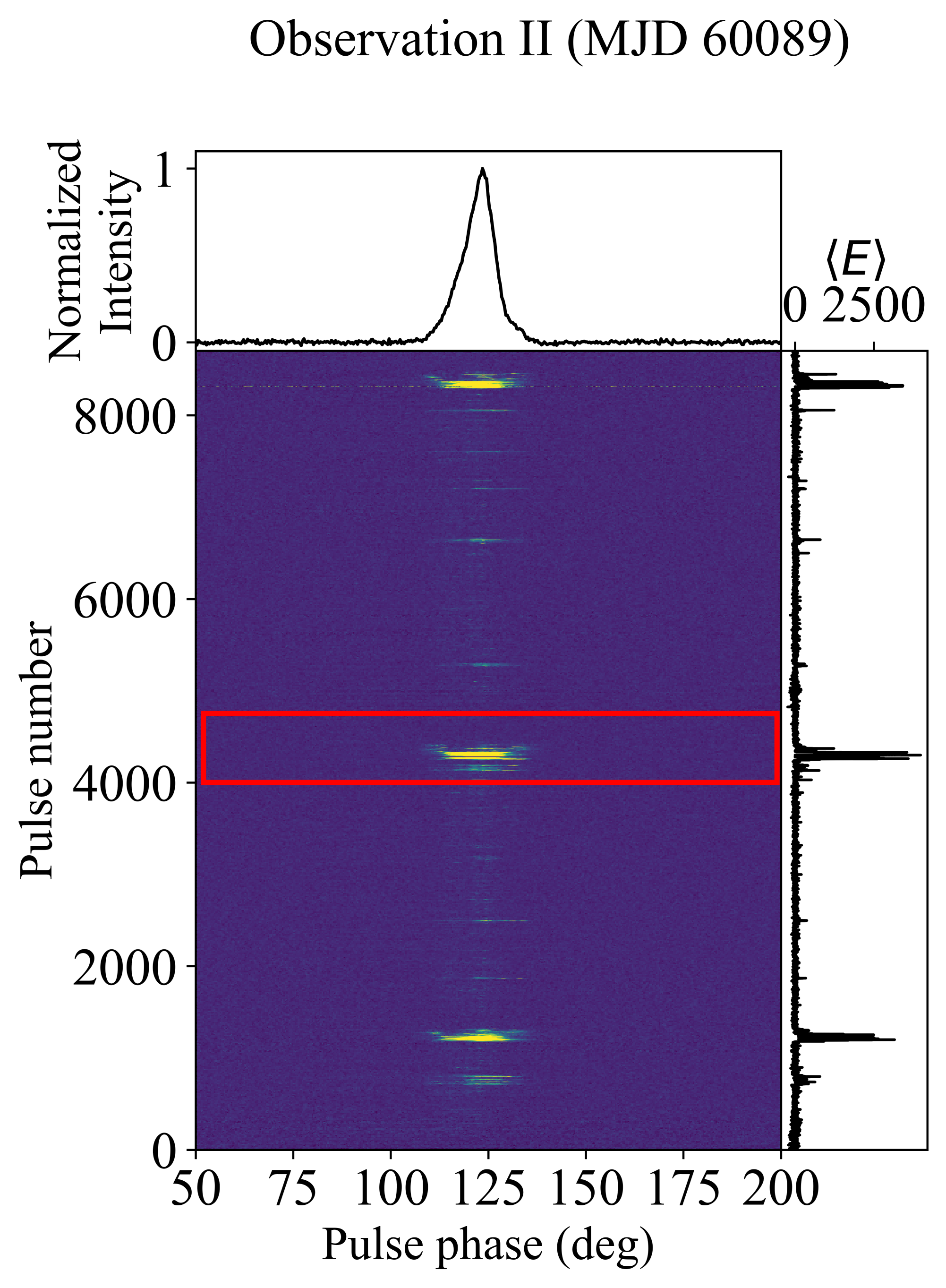}
        \caption{}
        \label{fig:imshow II}
    \end{subfigure}
    \hfill
    \begin{subfigure}{0.28\textwidth}
    \centering
        \includegraphics[width=\columnwidth,height=0.6\textheight,keepaspectratio]{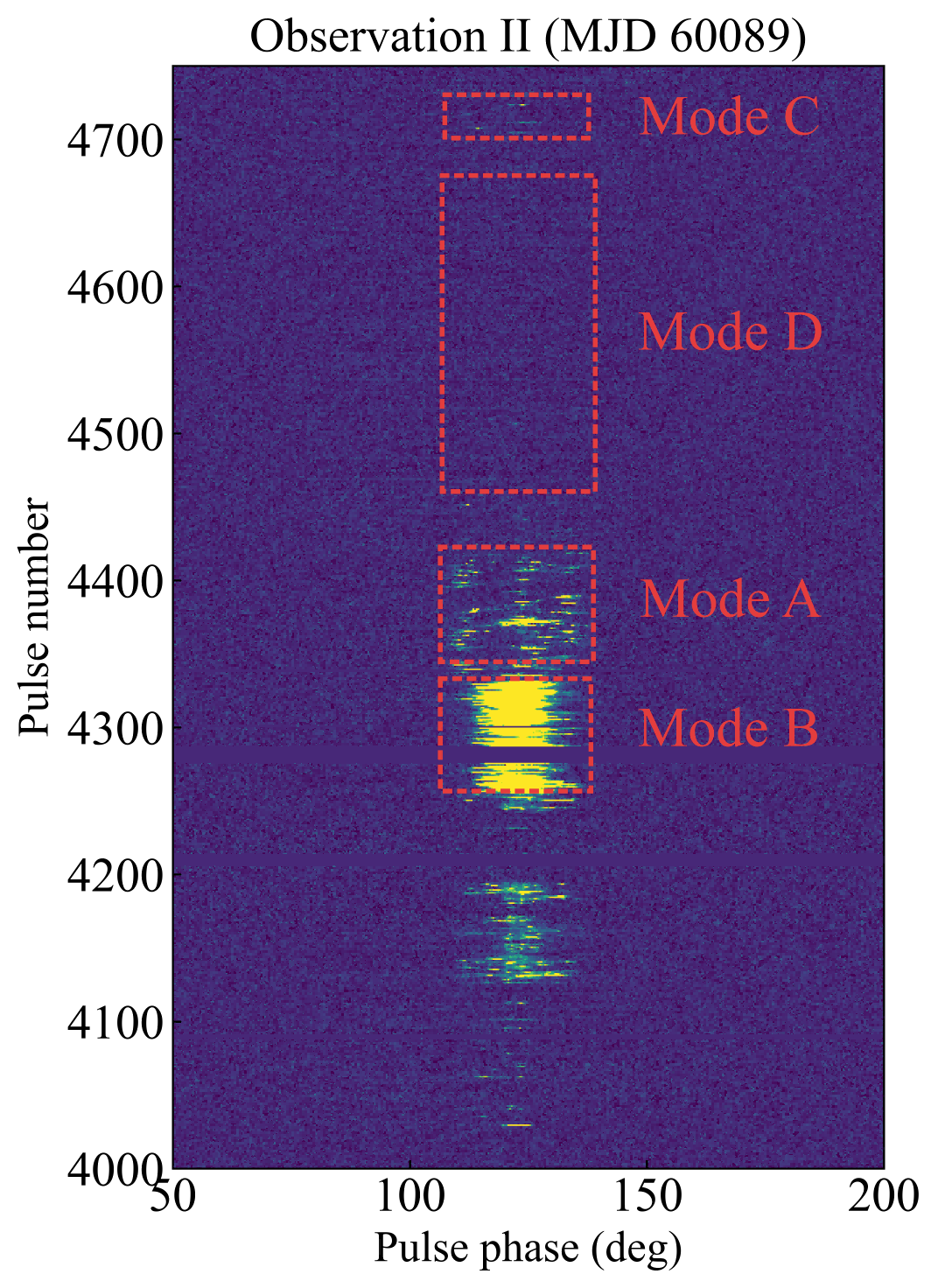}
        \caption{}
        \label{fig:imshow II_4000}
    \end{subfigure}

    \caption{Based on two independent observations, the single-pulse energy properties of 
PSR~J1820$-$0509 are illustrated. The central panels show stacked sequences 
of consecutive pulses; the upper panels present the integrated pulse profiles 
constructed from all pulses; and the right panels display the single-pulse 
energy distributions. It is evident that the single-pulse energies exhibit a 
multi-peaked distribution rather than a smooth continuous variation.The left 
set of panels corresponds to the first observation, the middle set to the 
second observation, and the right panel shows an enlarged view of the region marked by the red solid line in the middle panel (pulse numbers 4000–4750), including bright, weak, and null emission states.This clearly demonstrates the multi-peaked nature of the single-pulse energy distribution, rather than 
a uniform or continuous variation.}
    \label{fig:imshow} 
\end{figure*}

Figure~\ref{fig:imshow} presents the single-pulse behavior of PSR~J1820$-$0509 in two observing sessions. 
It is evident that the emission does not follow a single continuous variation process, 
but instead shows a clear state-separation behavior. Some pulses appear as relatively 
strong single-peaked bursts, some as weaker but longer-lasting enhanced emission episodes, 
while others occur only as short-duration isolated burst events. In addition, a large 
fraction of pulses remain in an extremely weak or undetectable low-energy state. 
The enlarged pulse sequence (see Figure~\ref{fig:imshow II_4000}) further shows that these different energy 
levels occur in clusters in time, demonstrating that the single-pulse energy distribution 
of this pulsar intrinsically exhibits multiple emission forms. Based on the properties 
of the pulses in the time series and the corresponding morphology of the averaged pulse 
profiles, we identify and define four emission states in the single-pulse sequence:
(1) Mode~A: relatively weak multi-component emission;
(2) Mode~B: stronger single-component emission;
(3) Mode~C: short-duration single-pulse burst events that appear in isolation within 
the time series;
(4) Mode~D: a state in which the emission energy of individual pulses falls below the 
detection threshold, while the averaged profile still retains a significant 
signal-to-noise ratio.

\begin{figure}
    \centering
    \includegraphics[width=\columnwidth,height=0.6\textheight,keepaspectratio]{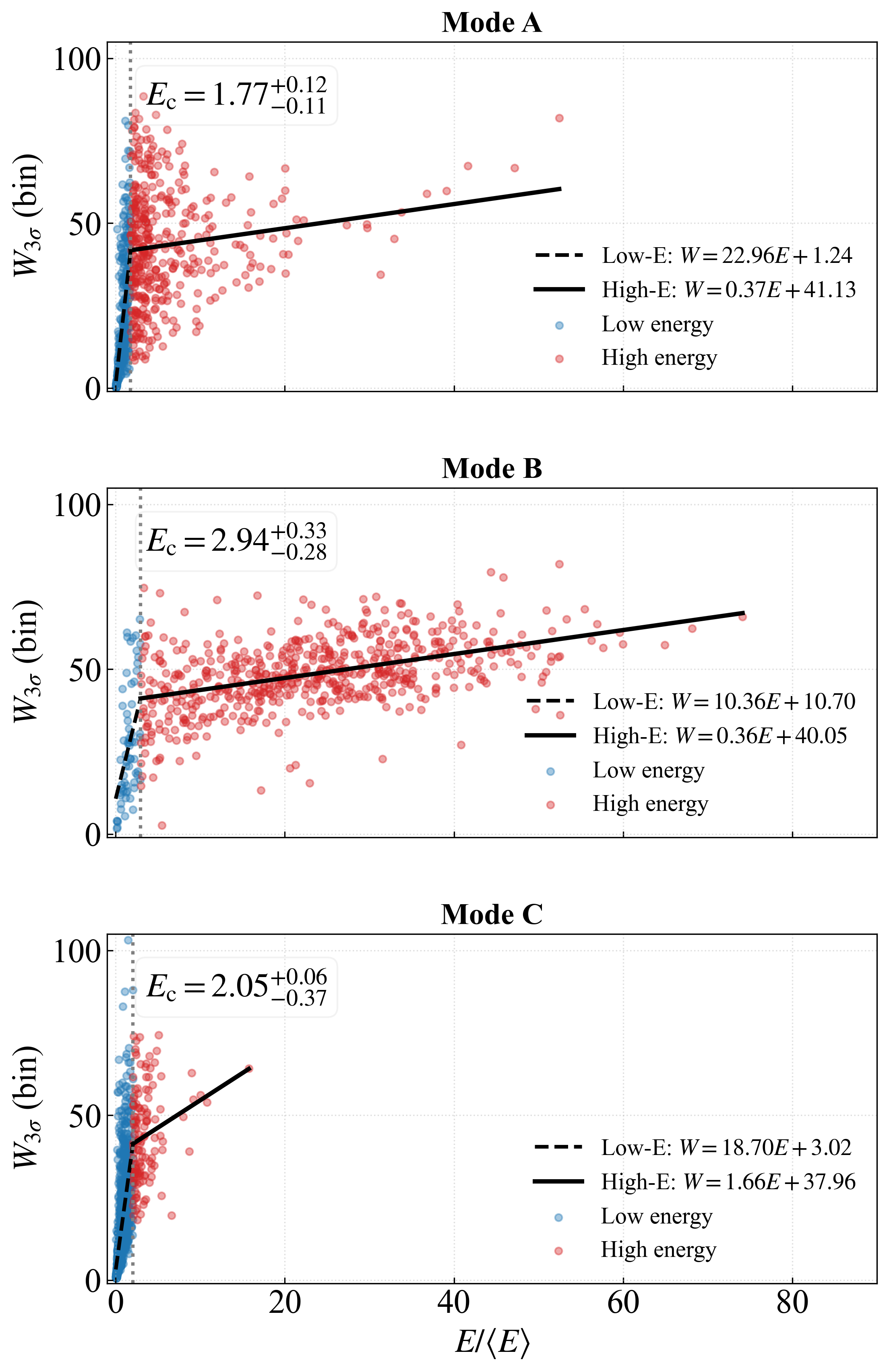}
    \caption{Relationship between the relative pulse energy and pulse width ($W_{3\sigma}$) for different classes of single pulses. 
The horizontal axis shows the pulse energy normalized by the mean energy of all pulses, $E/\langle E \rangle$. 
From top to bottom, the panels correspond to mode~A, mode~B, and mode~C. 
The blue and red points represent pulses classified as the weak emission state and the strong emission state, respectively. 
The black dashed and solid lines denote the linear fits in the low-energy and high-energy regimes, respectively. 
The vertical dotted line marks the classification threshold $E_{\mathrm{cross}}$, defined as the intersection of the two fitted relations.
}
    \label{fig:energy-With}
\end{figure}

Figure~\ref{fig:energy-With} shows the distributions of pulse width $W_{3\sigma}$ versus normalized 
energy $E/\langle E \rangle$ for the three emission modes. The three modes differ 
clearly in their distribution ranges, density concentrations, and scatter 
properties in the $W_{3\sigma}$--$E/\langle E \rangle$ plane. 
Mode~A is mainly concentrated in the low-energy region, with a broad range of 
$W_{3\sigma}$ values and a fraction of points extending to higher energies. 
Mode~B covers the widest energy range and shows the most complete overall 
distribution, with $W_{3\sigma}$ increasing with $E/\langle E \rangle$ and 
exhibiting a relatively clear positive trend. Mode~C is likewise concentrated 
in the low-energy region, with only a small number of points extending to higher 
energies and relatively large scatter in $W_{3\sigma}$.

Since the width--energy relations show different slopes in the low- and 
high-energy regimes, they cannot be adequately described by a single linear 
function. We therefore fitted the three modes separately using a continuous 
broken-line model,
\begin{equation}
W(E) = a + b_1 E + b_2 \max(0,\,E - E_{\rm c}),
\end{equation}
in which the transition energy $E_{\rm c}$ was treated as a free parameter and 
determined through a grid search. Specifically, for each mode, the 
$E/\langle E \rangle$ values were sorted in ascending order, and candidate 
$E_{\rm c}$ values were uniformly sampled over the central energy range after 
excluding the lowest and highest 10\% of the data. A broken-line fit was then 
performed for each candidate $E_{\rm c}$, and the value giving the minimum 
residual sum of squares (RSS) was adopted as the best-fitting transition energy. 
Parameter uncertainties were estimated using bootstrap resampling.
The best-fitting transition energies are 
$1.77^{+0.12}_{-0.11}$, $2.94^{+0.33}_{-0.28}$, and $2.05^{+0.06}_{-0.37}$ for 
Modes~A, B, and C, respectively. Among them, Mode~B has the highest $E_{\rm c}$, 
indicating that its rapid width growth extends over the broadest energy range, 
while Mode~A shows the steepest increase in the low-energy regime, indicating 
the strongest sensitivity of width to energy variations at low energies. 
Mode~C also shows a clear transition; however, because the high-energy branch 
contains only a small number of data points, the post-break slope is not well 
constrained and should be regarded as descriptive only, rather than used for 
further quantitative comparison.

\begin{figure*} 
    \centering
    \begin{subfigure}{0.6\textwidth}
        \centering
        \includegraphics[width=\linewidth]{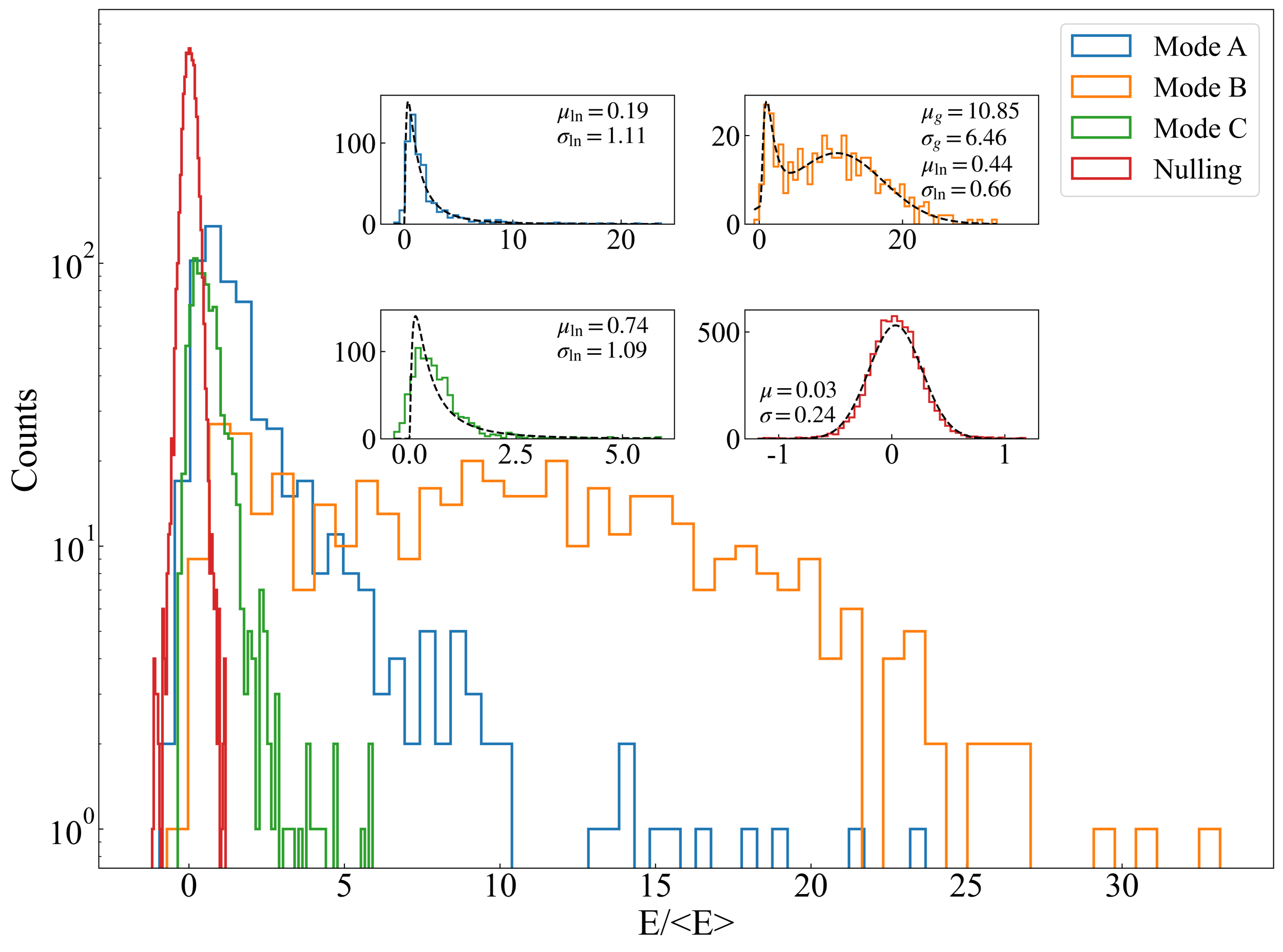} 
        \caption{}
        \label{fig:modes_c}
    \end{subfigure}
    \hfill
   \begin{subfigure}{0.6\textwidth}
        \centering
        \includegraphics[width=\linewidth]{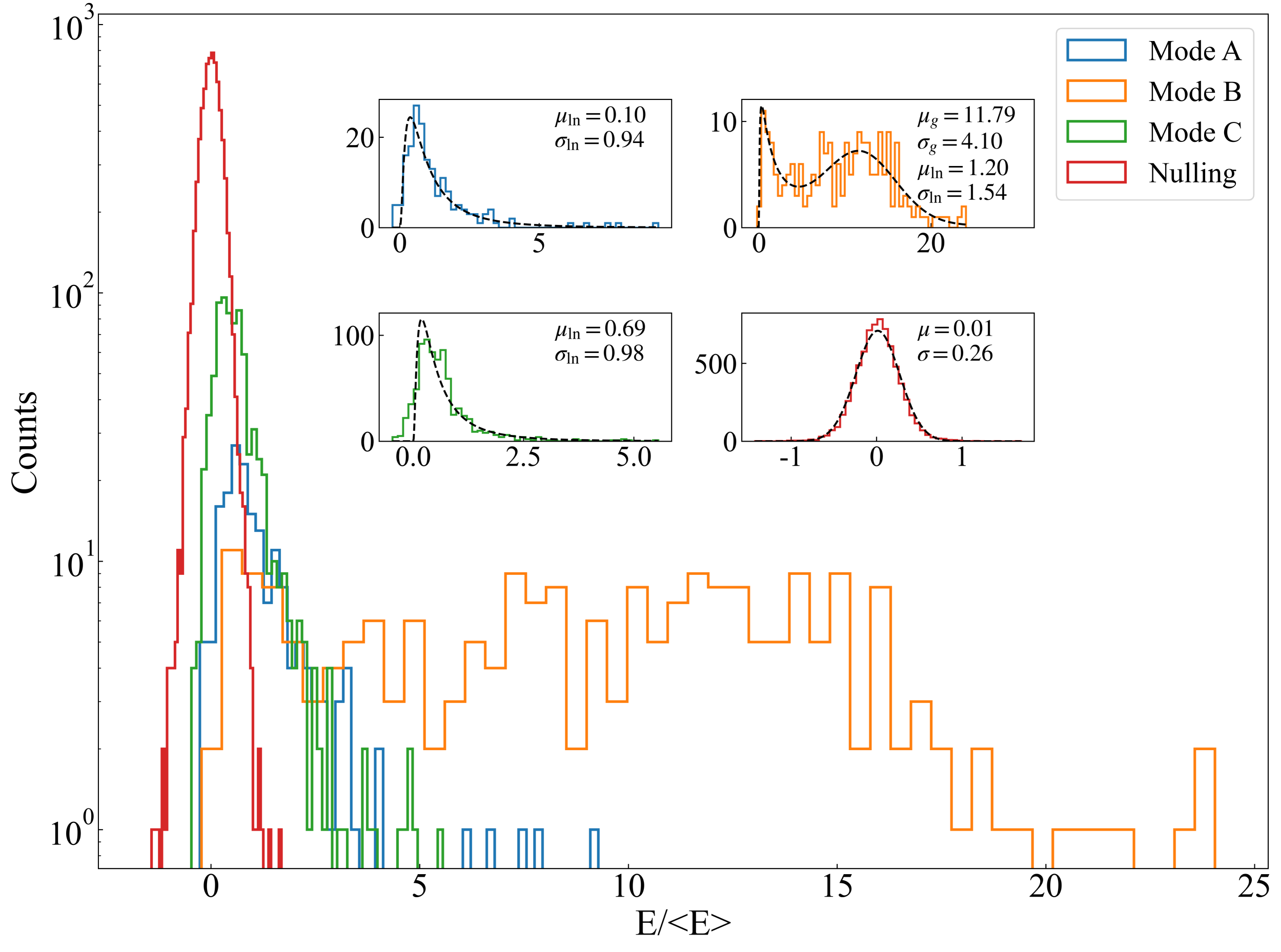}   
        \caption{}
        \label{fig:modes_d}
    \end{subfigure}
    \caption{Histograms of the relative energy distributions for different emission modes. The horizontal axis shows the normalized pulse energy $E/\langle E \rangle$, and the vertical axis gives the number of pulses in each energy bin (logarithmic scale). Mode~A (blue), Mode~B (orange), Mode~C (green), and the nulling state (red) are shown for comparison. The upper and lower panels correspond to the first and second observations, respectively. The distributions reveal a clear four-component structure associated with the different emission states.
}
    \label{fig:energy distribution}
\end{figure*}

 To further investigate the single-pulse energy properties of PSR J1820$-$0509, we present the pulse energy distribution (Figure~\ref{fig:energy distribution}).
This picture presents the relative energy distributions for the four emission modes. Considering that the occurrence fractions of the four emission states are not identical in the two observing epochs, we separately display the energy distributions for each state in the two datasets to avoid potential bias in the overall distribution shape arising from differences in mode occupancy.
A comparison between the two epochs shows that, within statistical uncertainties, no significant systematic differences are observed in the integrated energy distributions of the individual emission states. This indicates that the statistical properties of the pulsar's emission energy remain stable across the two observations.
To further examine the statistical behavior of the different emission states, we performed fits to the single-pulse normalized energy distributions ($E/\langle E\rangle$) for each epoch independently. An identical fitting procedure was applied to both datasets. The resulting functional forms are highly consistent between the two epochs, with only minor differences in the fitted parameters.

In both observations, Mode~A follows a lognormal distribution in the two observing epochs, with parameters $(\mu_{\ln}=0.19, \sigma_{\ln}=1.11)$ for the first observation and $(\mu_{\ln}=0.10, \sigma_{\ln}=0.94)$ for the second. Similarly, Mode~C also exhibits a lognormal distribution, with $(\mu_{\ln}=0.74, \sigma_{\ln}=1.09)$ and $(\mu_{\ln}=0.69, \sigma_{\ln}=0.98)$ in the two epochs, respectively. Compared with Mode~A, Mode~C has a larger logarithmic mean, while both modes display characteristic long-tailed distributions. The stability of these parameters across epochs indicates good reproducibility of the underlying statistical behavior. Lognormal statistics naturally arise from multiplicative stochastic processes and have been widely reported in pulsar single-pulse emission \citep{cairns2001intrinsic,burke2012high}, as well as in accreting systems \citep{uttley2005non}, suggesting that the emission of Mode~A and Mode~C may originate from multiplicative plasma fluctuation processes within the pulsar magnetosphere. In contrast, Mode~B exhibits a clear double-peaked structure in both observing sessions, which can be described by the superposition of a Gaussian and a lognormal component. For the first observation, the Gaussian component has $(\mu_{\mathrm{g}} = 10.85,\, \sigma_{\mathrm{g}} = 6.46)$ and the lognormal component has $(\mu_{\ln} = 0.44,\, \sigma_{\ln} = 0.66)$, while for the second observation the corresponding parameters are $(\mu_{\mathrm{g}} = 11.79,\, \sigma_{\mathrm{g}} = 4.10)$ and $(\mu_{\ln} = 1.20,\, \sigma_{\ln} = 1.54)$. Although these parameters vary slightly between epochs, the double-peaked structure remains statistically significant, with the Gaussian component dominating the high-energy part and the lognormal component contributing to the extended low-energy tail. This suggests that Mode~B may represent the superposition of two distinct emission processes: a relatively stable high-energy component and another driven by multiplicative fluctuations. Finally, Mode~D, corresponding to the null (pseudo-nulling) state, shows an energy distribution strongly concentrated near zero and well described by a Gaussian distribution, with $(\mu=0.03, \sigma=0.24)$ and $(\mu=0.01, \sigma=0.26)$ in the two observations. The near-zero mean and symmetric shape indicate that this state is dominated by system noise, consistent with noise statistics \citep{wang2007pulsar}.


\subsection{Mode Change }


Based on the mode classification results in Section 3.2, we further counted the number of pulses in Mode A, Mode B, and Mode C.In the first observing session, the numbers of pulses identified as Mode A, Mode B, and Mode C are 577, 443, and 888, respectively. In the second observing session, the corresponding numbers are 184, 219, and 802.
Following the methods of \citet{rahaman2021mode} and \citet{basu2023single}, we visually identified the emission mode of each individual pulse and grouped consecutive pulses with the same mode into continuous segments. The duration of each mode segment is defined as the number of consecutive pulses belonging to that mode and is expressed in units of the pulsar rotation period $P$. Based on this definition, the durations of all segments corresponding to the same emission mode were averaged to obtain the mean duration of that mode (see Table~\ref{tab:2}). The relative abundance is defined as the fraction of pulses belonging to a given mode with respect to the total number of pulses, including null pulses.

\begin{table}
    \centering
    \caption{Statistical results of different emission states in the two observing sessions. The first column lists the observing date and the corresponding Modified Julian Date (MJD). Columns 2--5 present the emission mode, the number of pulses ($N_{\rm p}$), the relative abundance with respect to the total number of pulses, and the average duration of consecutive segments for that mode (in units of the rotation period $P$), respectively. The statistics for the two observing sessions are calculated independently.}
    \label{tab:2}
    \resizebox{0.5\textwidth}{!}{ 
    \begin{tabular}{|c|c|c|c|c|}
        \hline
        Observation time (MJD) & Mode & Np & Relative abundance (\%) & Average duration (P) \\
        \hline
        \multirow{3}{*}{Observation I (MJD 59576)} & A & 577 & 6.74 & 66.00 \\
  
        & B & 443 & 5.18 & 75.33 \\

        & C & 888 & 10.38 & 1.76 \\
        
        \hline
        \multirow{3}{*}{Observation II (MJD 60089)} & A & 184 & 2.16 & 45.00 \\

        & B & 219 & 2.57 & 74.00 \\
  
        & C & 802 & 9.41 & 1.74 \\
        \hline
    \end{tabular}
    }
\end{table}

\begin{figure*} 
    \centering

    \begin{subfigure}{0.45\textwidth}
        \centering
        \includegraphics[width=\linewidth]{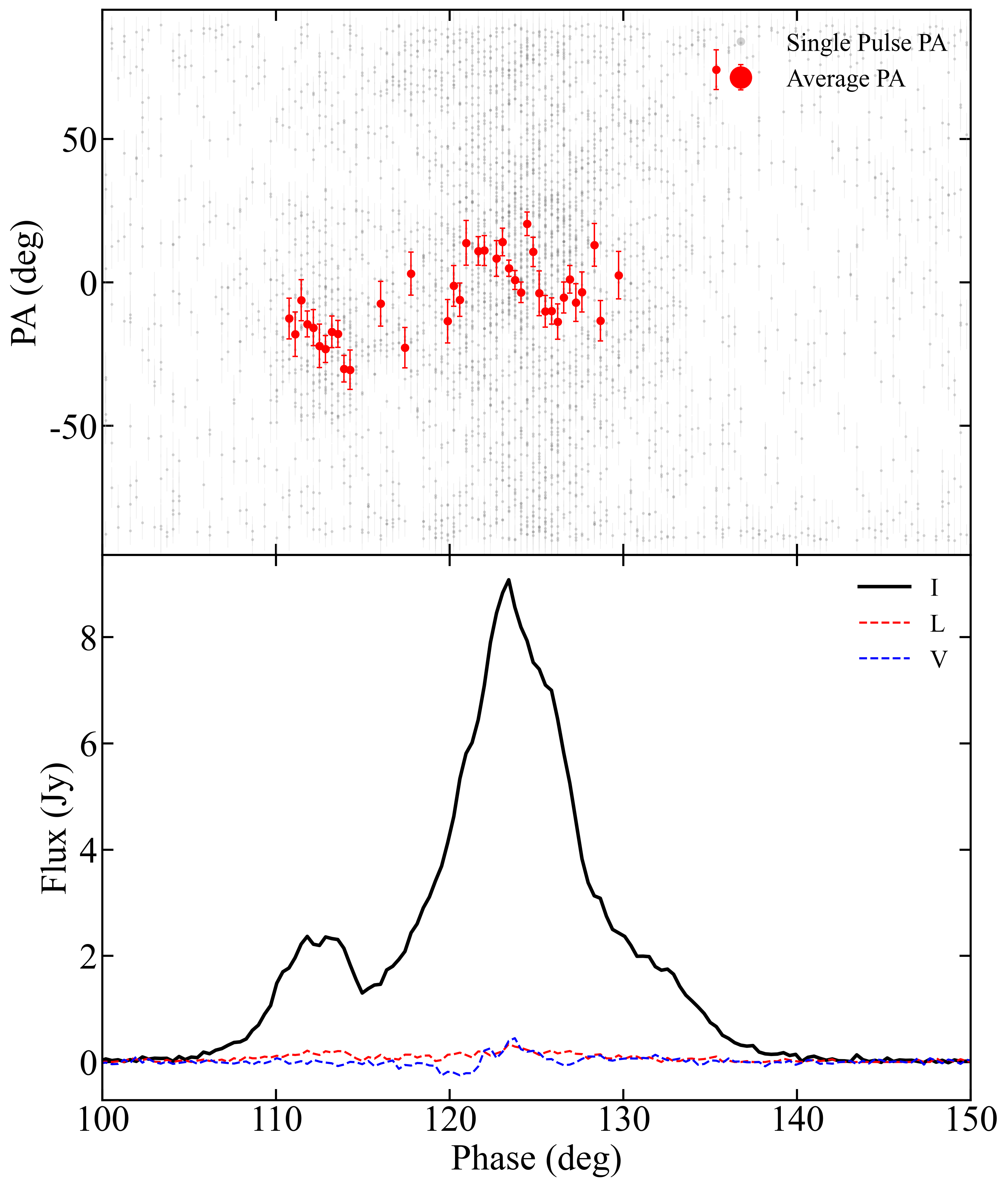}
        \caption{}
        \label{fig:modes_a}
    \end{subfigure}
    \hfill 
    \begin{subfigure}{0.45\textwidth}
        \centering
        \includegraphics[width=\linewidth]{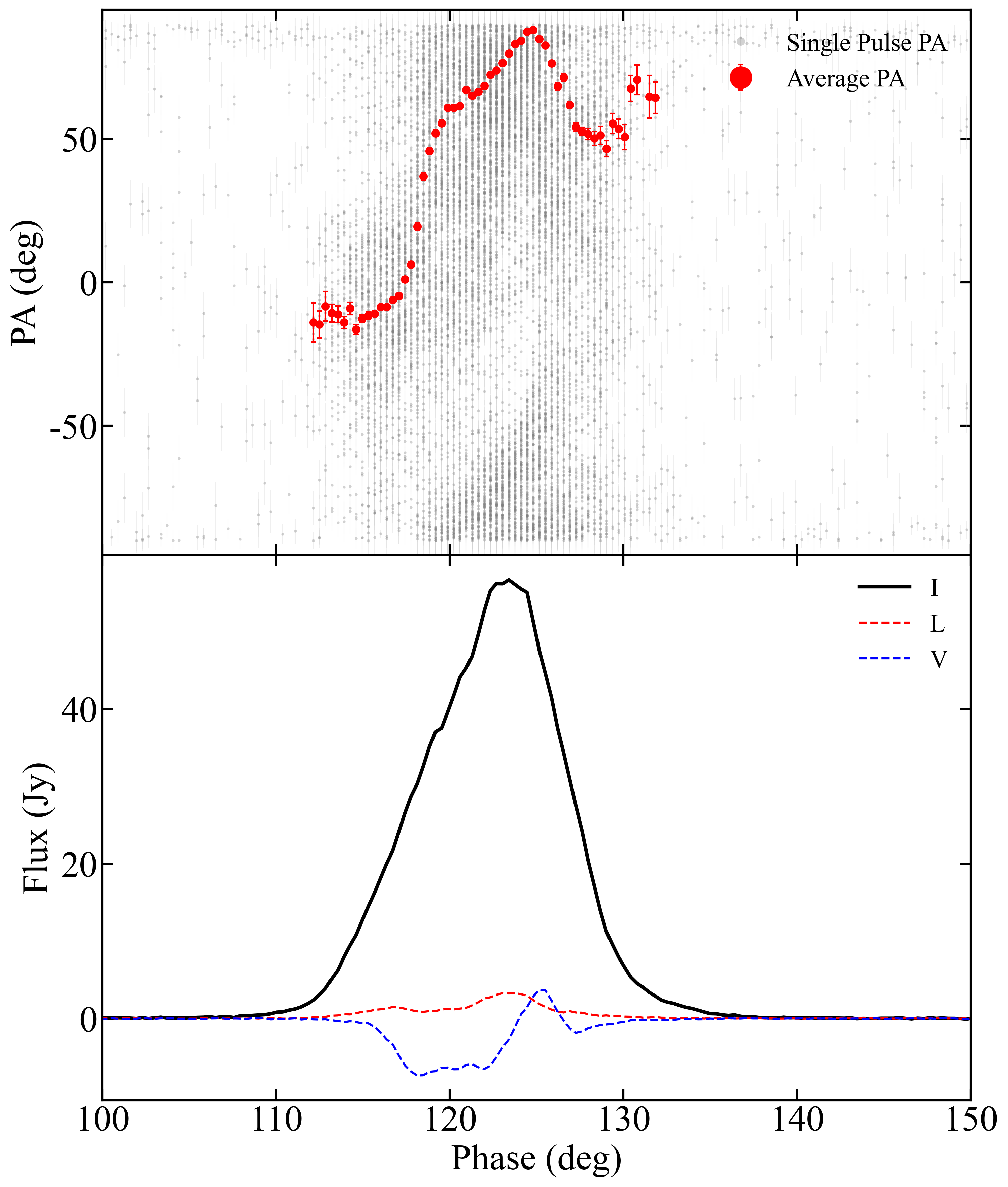}
        \caption{}
        \label{fig:modes_b}
    \end{subfigure}

    \vspace{1.5ex} 

    \begin{subfigure}{0.45\textwidth}
        \centering
        \includegraphics[width=\linewidth]{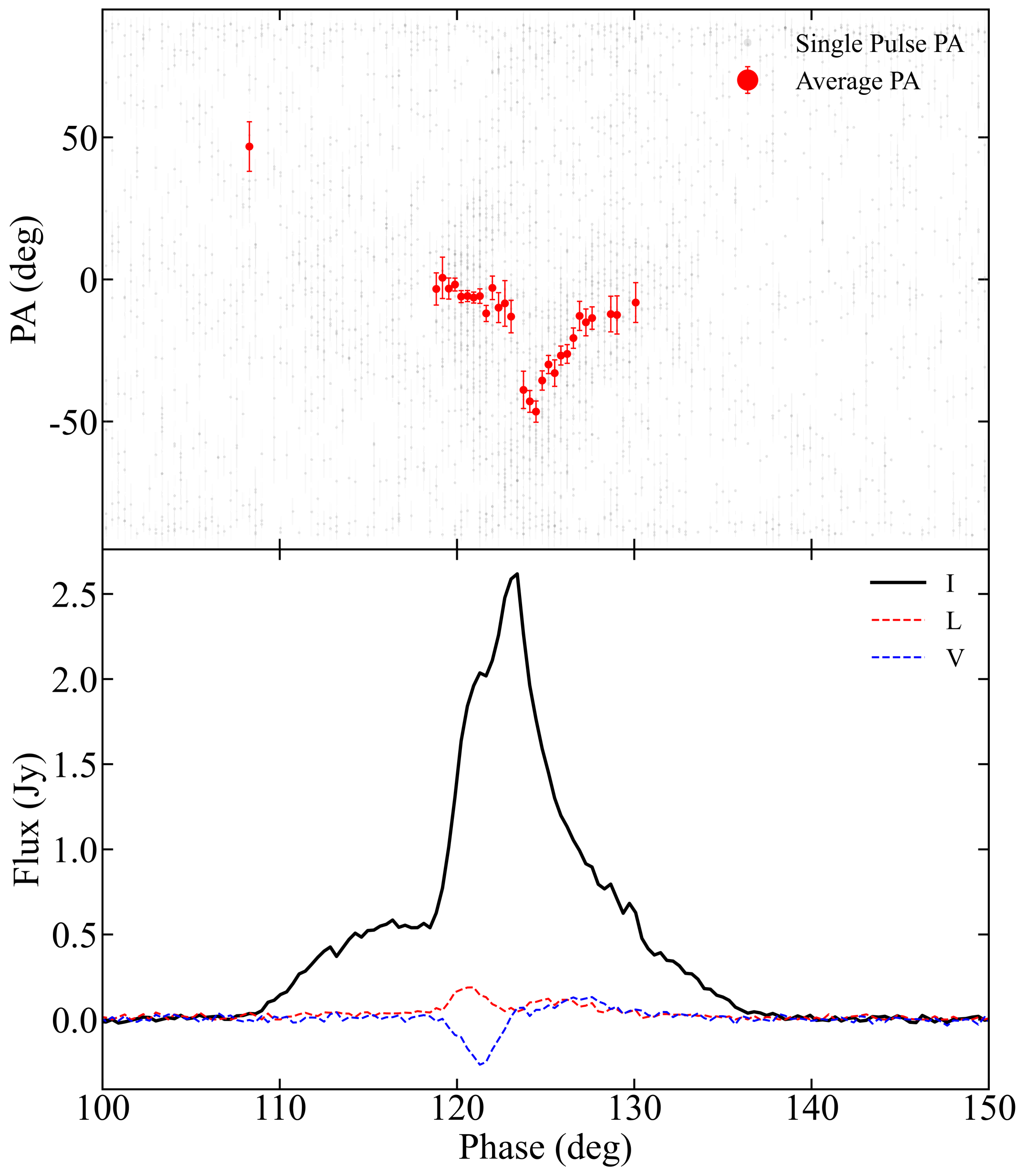} 
        \caption{}
        \label{fig:modes_c}
    \end{subfigure}
    \hfill
   \begin{subfigure}{0.45\textwidth}
        \centering
        \includegraphics[width=\linewidth]{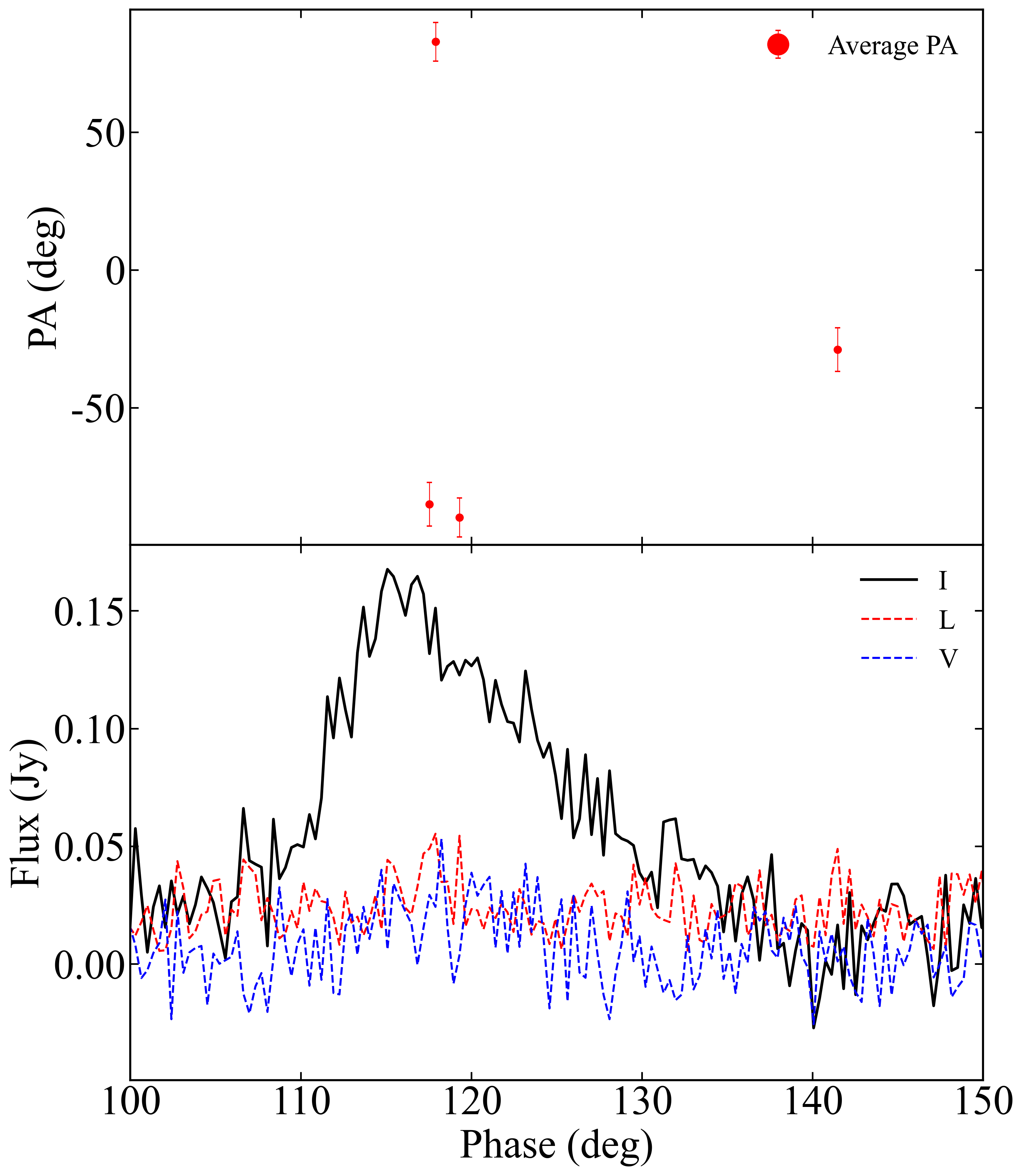}
        \caption{}
        \label{fig:modes_d}
    \end{subfigure}
    
    \caption{
Polarization properties of the four emission modes are presented. The upper 
panels show the polarization position angle (PA). In the lower main panels, 
the black, red, and blue curves represent the total intensity ($I$), linear 
polarization ($L$), and circular polarization ($V$) profiles, respectively. 
The four sub-panels (a--d) correspond to the following emission modes: 
Mode~A — relatively weak, multi-component emission; 
Mode~B — stronger, single-component emission; 
Mode~C — short-duration, isolated burst events; 
and Mode~D — single pulses with emission energies below the detection 
threshold, while the stacked average profile still exhibits a significant 
signal-to-noise ratio.
    }
    \label{fig:four_modes}
\end{figure*}

Figure~\ref{fig:four_modes} shows the total intensity, polarization properties, and position angle distributions obtained by stacking the four selected modes, which exhibit significant differences in these characteristics. In Mode~A, the overall pulse profile shows relatively low intensity and a moderately broadened structure. The circular polarization component is predominantly weakly negative across the pulse window, without a clear sense reversal. In the leading phase range ($110^\circ$--$120^\circ$), the negative circular polarization has a small amplitude and gradually weakens toward the main peak, with only an extremely weak positive component occasionally appearing at the trailing edge. The correlation between circular polarization and total intensity is therefore reduced. The extrema of circular polarization remain approximately aligned with the center of the main component, suggesting no significant shift in the emission geometry center, although the handedness structure becomes indistinct. Meanwhile, the linear polarization is generally weak and does not increase with total intensity, and the polarization position angle (PPA) within $-50^\circ$ to $25^\circ$ is highly scattered without systematic variation, indicating a strongly perturbed magnetospheric plasma environment with reduced coherence and geometric ordering \citep{jones2011instabilities,wang2007pulsar}. In contrast, Mode~B exhibits a more concentrated single-peaked profile with significantly enhanced total intensity. The circular polarization shows a clear and strong sense reversal: it is distinctly negative in the leading region (110$^\circ$--120$^\circ$), then rapidly increases and reverses to positive near the main peak (122$^\circ$--124$^\circ$), with amplitude positively correlated with intensity, and subsequently decreases together with total intensity beyond the peak. This ordered behavior reflects a spatial transition between emission regions with different magnetic or plasma conditions \citep{radhakrishnan1969magnetic}. The linear polarization increases with total intensity, and the PPA exhibits a smooth, monotonic swing (approximately $-30^\circ$ to $50^\circ$), consistent with the Rotating Vector Model (RVM), indicating a more stable and geometrically ordered magnetosphere \citep{radhakrishnan1969magnetic,rankin1983toward,rankin1993toward}. Mode~C, shown in Figure~\ref{fig:modes_c}, displays a clearly asymmetric total intensity profile with a broad leading component and a sharp main peak at $\sim122^\circ$. The fractional linear polarization remains low across the pulse window, although a localized enhancement appears just before the main peak. In contrast, the circular polarization reaches a strong negative extremum at the main peak, with $|V| > L$. The PPA deviates significantly from the classical RVM prediction, lacking a smooth S-shaped swing and instead showing a steep downward dip ($\sim122^\circ$--$124^\circ$) to about $-50^\circ$, coincident with the extrema of $I$ and $V$. This behavior is consistent with the superposition of unresolved orthogonal polarization modes (OPMs), which can suppress linear polarization \citep{stinebring1984pulsara,stinebring1984pulsarb}, and may also reflect core-component emission characteristics \citep{rankin1990toward} or propagation effects such as Faraday conversion that distort the intrinsic PA trajectory \citep{wang2010polarization}.
Finally, in Mode~D, the emission is almost completely suppressed, with the main peak intensity reduced to the noise level and only weak fluctuations remaining in the leading region. The circular polarization is confined to the leading region and remains persistently positive, while the linear polarization nearly vanishes and the PPA shows no discernible structure. This suggests that suppression of the main emission component removes the spatial differentiation of polarization, leaving only weak residual emission, possibly related to insufficient magnetospheric plasma supply or failure to meet stable emission conditions \citep{biggs1992analysis,gajjar2014frequency}.

\begin{figure}
    \centering
    \includegraphics[width=\columnwidth,height=0.6\textheight,keepaspectratio]{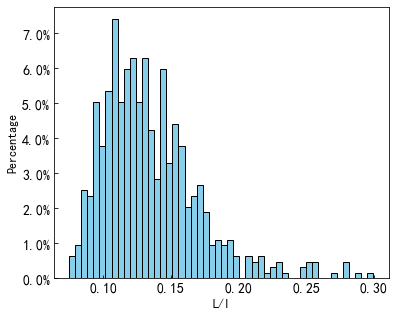} 
    \caption{The horizontal axis denotes the degree of linear polarization, and the vertical axis represents the fraction of counts in each bin.}
    \label{fig:V/I distribution}
\end{figure}

After systematically comparing the average polarization profiles and the corresponding PPA distributions of the four emission modes in the previous sections, we further investigate the statistical properties of polarization on the single-pulse timescale. Recent studies have shown that even for pulsars exhibiting complex PPA distributions, a systematic swing resembling the RVM may still be recovered when selecting subsamples with high polarization fractions (e.g., \cite{mitra2023evidence,johnston2024thousand}). This result suggests that the underlying magnetospheric geometry may remain preserved within strongly polarized pulse samples.
Motivated by this possibility, we examine the distribution of the linear polarization fraction for all single pulses (Figure~\ref{fig:V/I distribution}). The results show that although linear polarization is systematically present in all emission modes, the vast majority of pulses have polarization fractions below 0.3, and no highly polarized burst-like events are detected. In other words, on the single-pulse timescale the emission of this source is statistically dominated by low-to-moderate polarization levels. Such relatively low polarization fractions indicate that the burst emission is unlikely to be dominated by a single stable polarization mode. Instead, a more plausible explanation is that the observed depolarization arises from the incoherent superposition of OPMs. This polarization reduction caused by the mixture of two nearly orthogonal propagation modes has been widely observed and discussed in pulsar radio emission studies (e.g., \cite{manchester1975observations,smits2006frequency}).




\subsection{Periodicity of burst episode}

\begin{figure*}
    \centering
    \begin{subfigure}{1\columnwidth}
       \centering
        \includegraphics[width=\columnwidth,height=0.6\textheight,keepaspectratio]{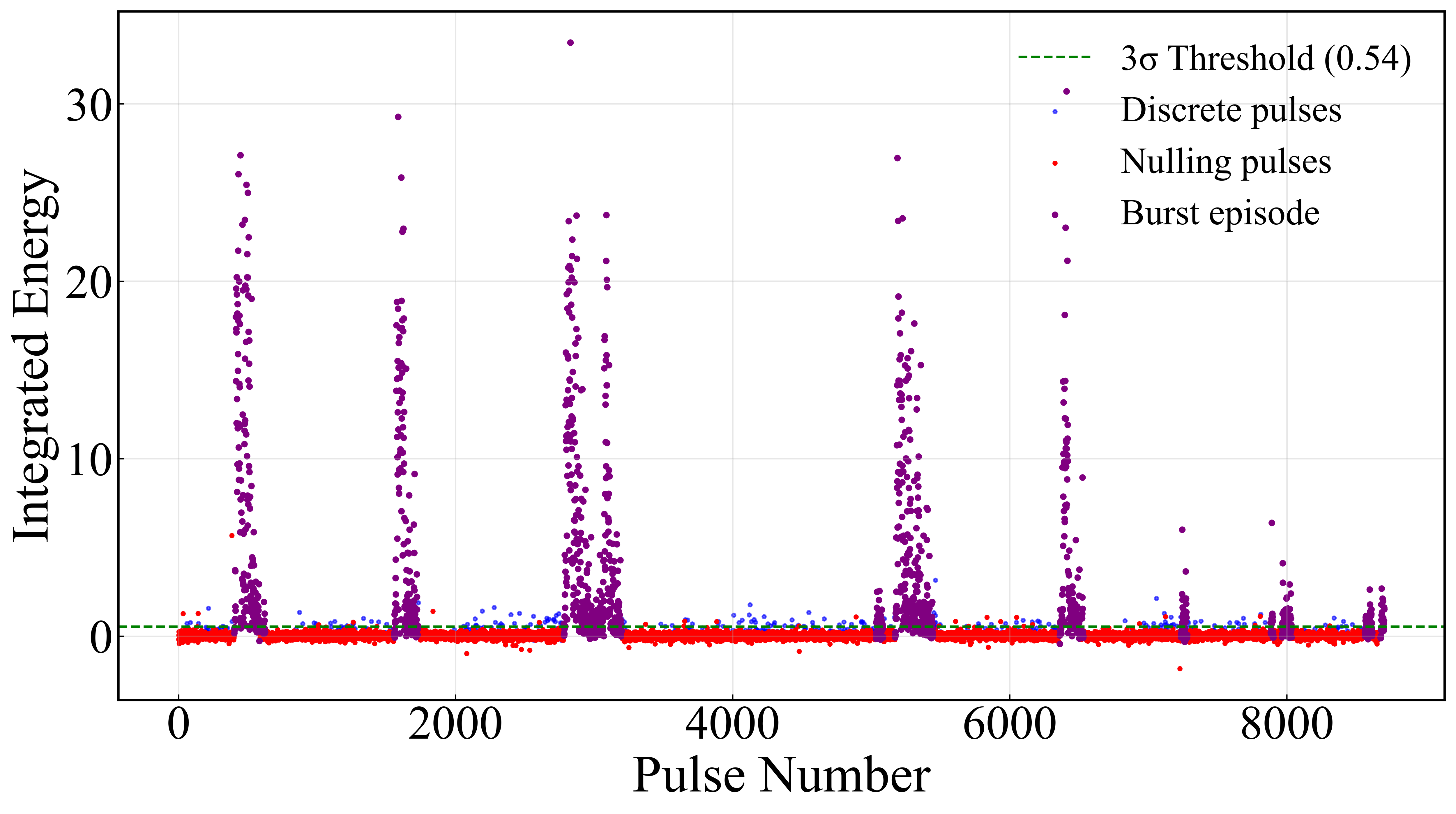}
        \caption{}
        \label{fig:burst 1}
    \end{subfigure}
    \vskip\baselineskip
    \begin{subfigure}{1\columnwidth}
        \centering
        \includegraphics[width=\columnwidth,height=0.6\textheight,keepaspectratio]{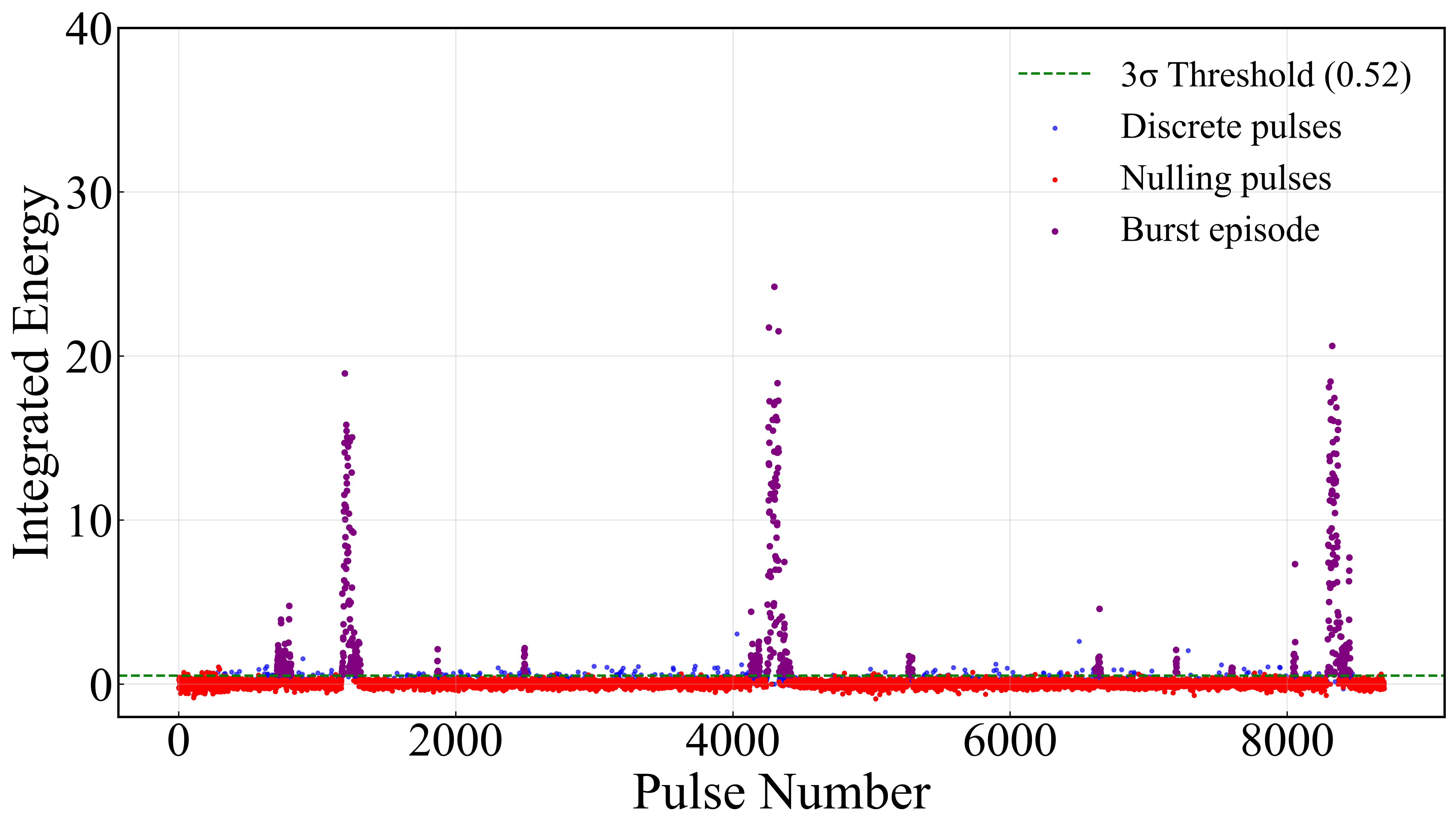}
        \caption{}
        \label{fig:burst 2}
    \end{subfigure}
    \caption{The method for identifying concentrated bursts is as follows. Red dots represent null pulses, blue dots denote normal pulses, and purple dots mark the identified concentrated burst pulses. The green dashed line corresponds to the $3\sigma$ threshold of the null-pulse energy distribution. (a) Results from the first observation; (b) Results from the second observation.
}
    \label{fig:concentrated_burst}
\end{figure*}

In our observations, significant burst emission were detected. The pulsar's radiation intensity significantly increased over several tens to hundreds of rotation periods, followed by a long period of null state. To quantitatively characterize this phenomenon, we used the following criterion: if the energies of five consecutive pulses exceed the $3\sigma$ threshold of the zero-pulse energy distribution, the interval is classified as a “burst cluster” (purple dots in Figure~\ref{fig:concentrated_burst}); the remaining pulses are classified as “isolated bursts” (yellow points in Figure~\ref{fig:concentrated_burst}). Based on the above criteria, the characteristics of burst episode in the two observations show significant differences. In the first observation, 1318 pulses exceeded the \colorbold{black}{$3\sigma_{\rm E,null}$} threshold, with 1125 pulses (approximately 85\%) meeting the criterion for burst episode. These burst episode pulses are mainly distributed within six intervals of strong bursts with relative energy $\ge10$ (Each burst episode spans from tens to hundreds of rotation periods). In the second observation, the number of pulses exceeding the $3\sigma_{\rm }$ threshold decreased to 751 (approximately 43\% less than the first observation), with 580 pulses (about 77\%) classified as burst episode. These pulses are primarily concentrated in three strong-burst intervals with relative energy $\ge10$, indicating a significant reduction compared to the first observation. The differences in the number of burst episode observed in the two sessions suggest that PSR~J1820$-$0509 exhibits two distinct states: ``frequent clusters'' and ``sparse clusters,'' with its burst activity showing significant fluctuations over time. This conclusion aligns with the observations made by N. Wang \citep{wang2007pulsar}.



\begin{figure*}
    \centering
    \begin{subfigure}{0.45\textwidth}
        \centering
        \includegraphics[width=\columnwidth,height=0.6\textheight,keepaspectratio]{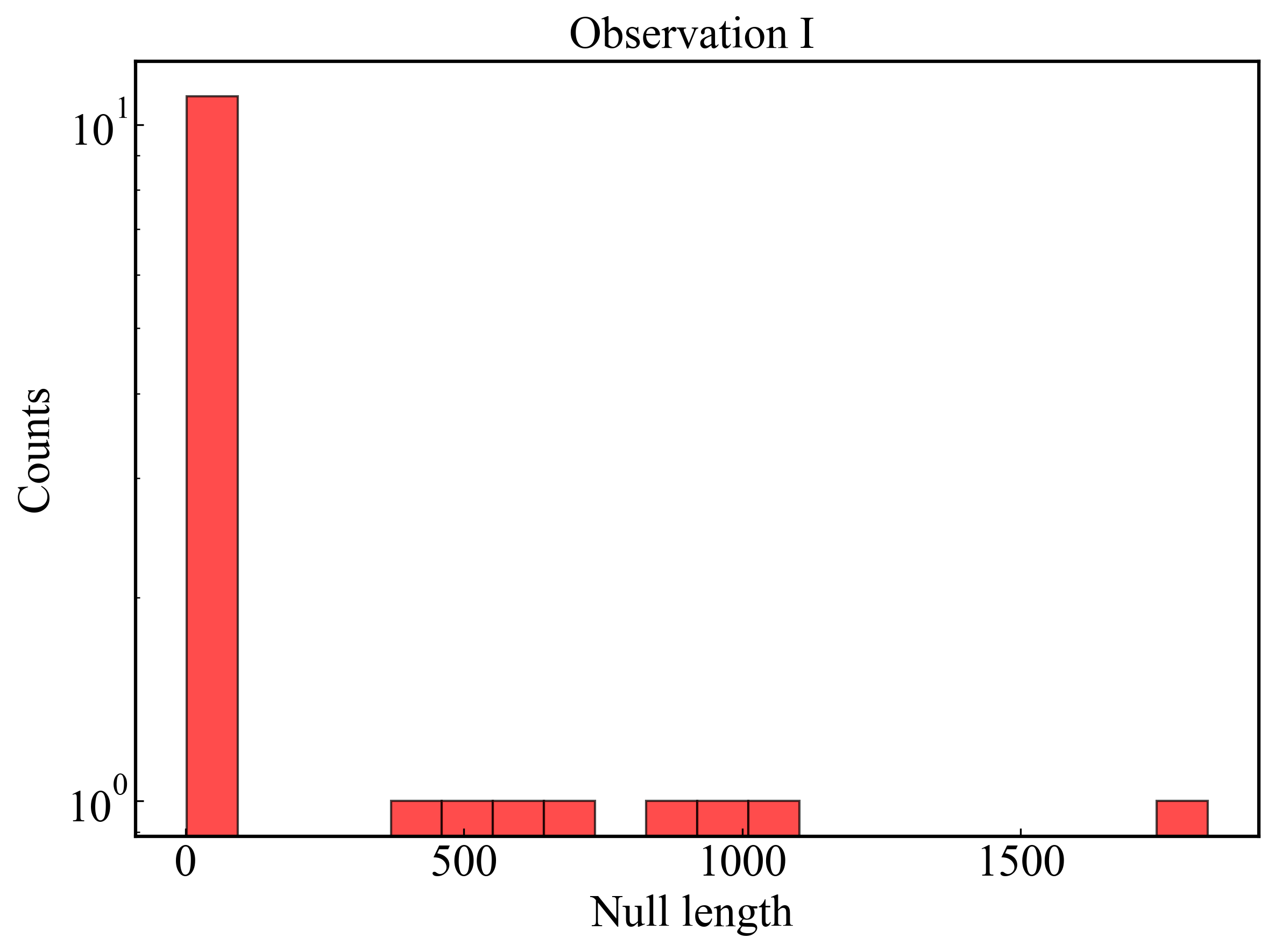}
        \caption{}        
        \label{fig:null I}
    \end{subfigure}\hfill
    \begin{subfigure}{0.45\textwidth}
        \centering
        \includegraphics[width=\columnwidth,height=0.6\textheight,keepaspectratio]{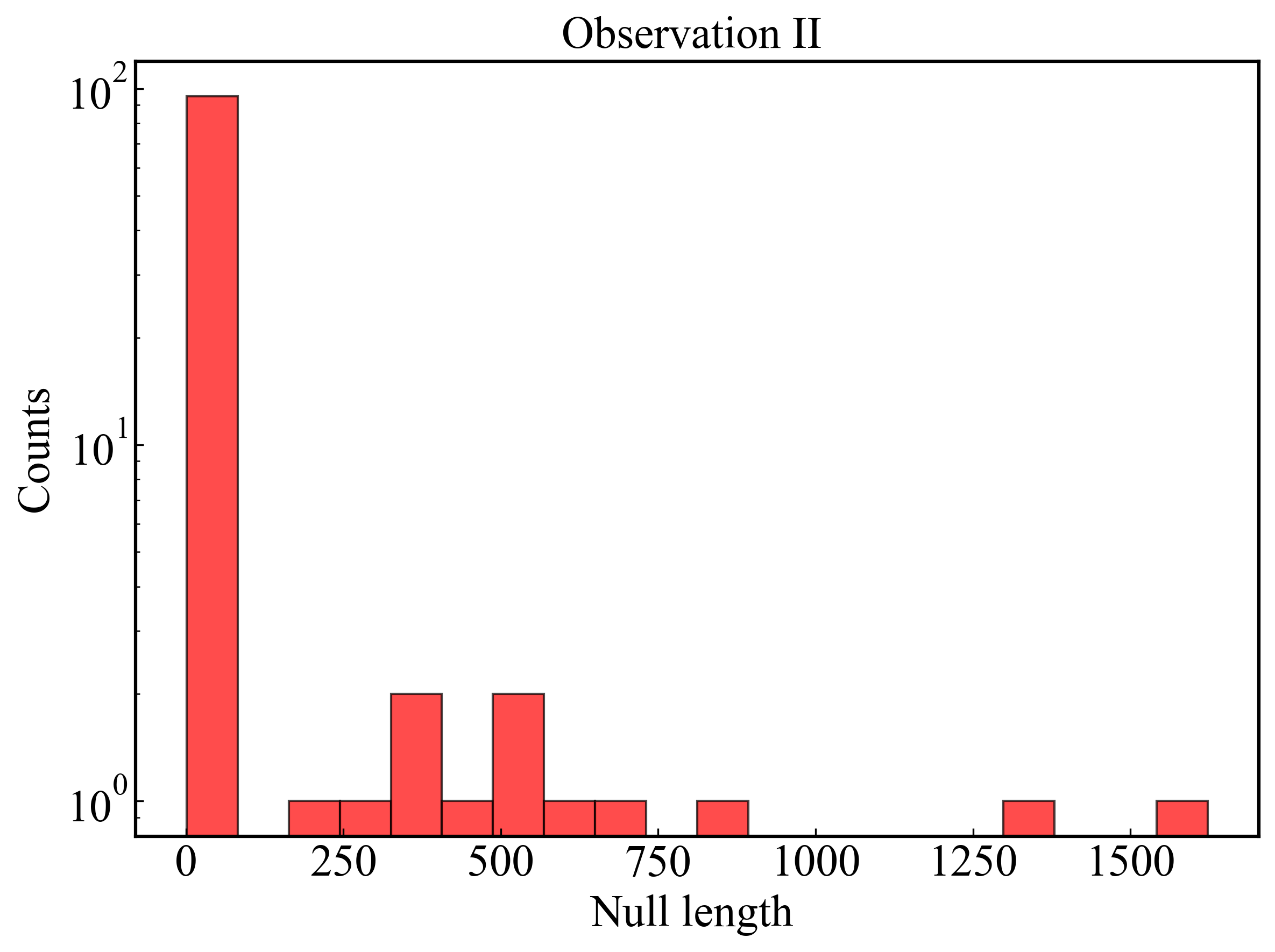}
        \caption{}        
        \label{fig:null II}
    \end{subfigure}\hfill
    \begin{subfigure}{0.45\textwidth}
        \centering
        \includegraphics[width=\columnwidth,height=0.6\textheight,keepaspectratio]{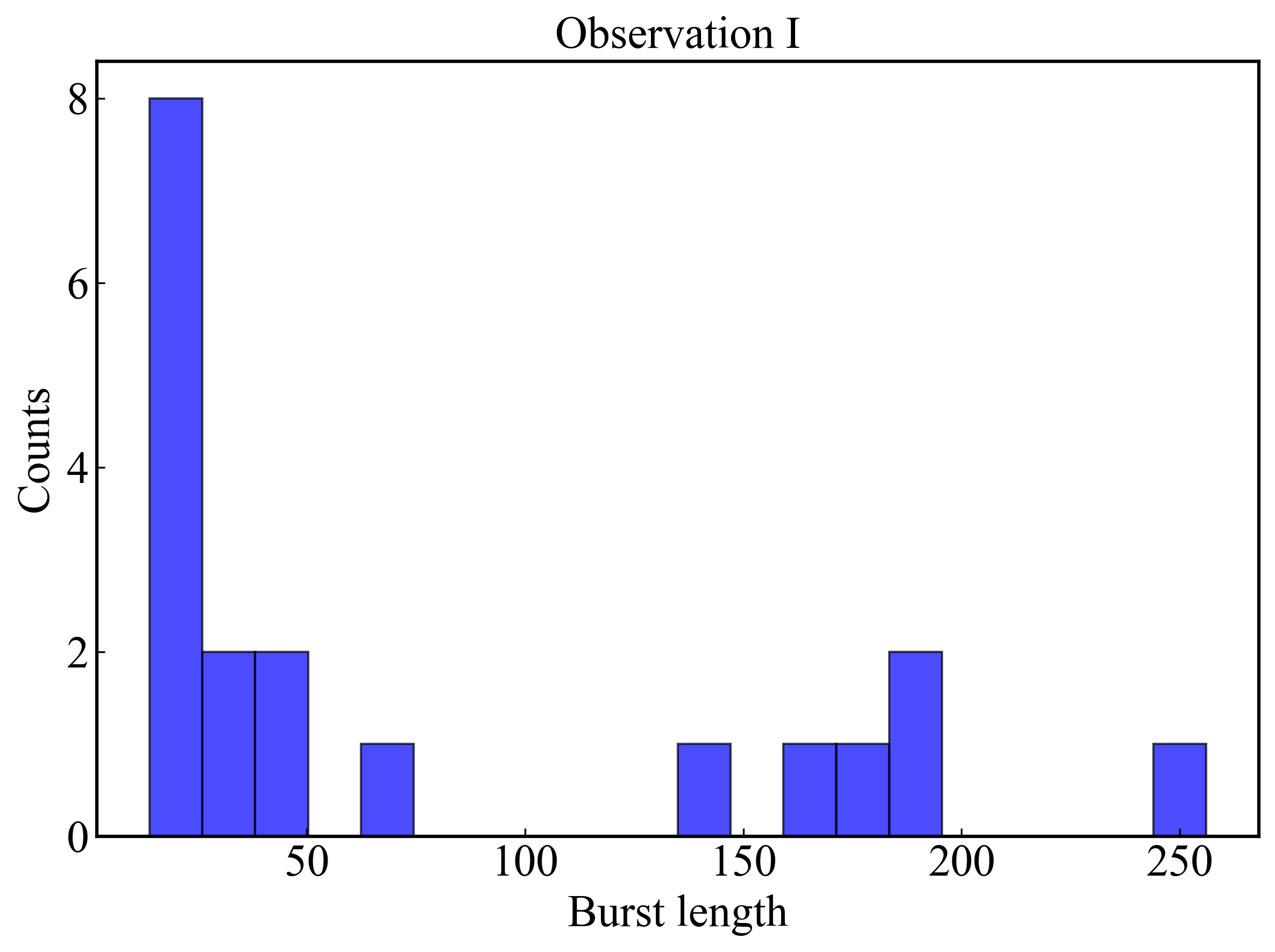}
        \caption{}
        \label{fig:burst I}

    \end{subfigure}\hfill
    \begin{subfigure}{0.45\textwidth}
        \centering
        \includegraphics[width=\columnwidth,height=0.6\textheight,keepaspectratio]{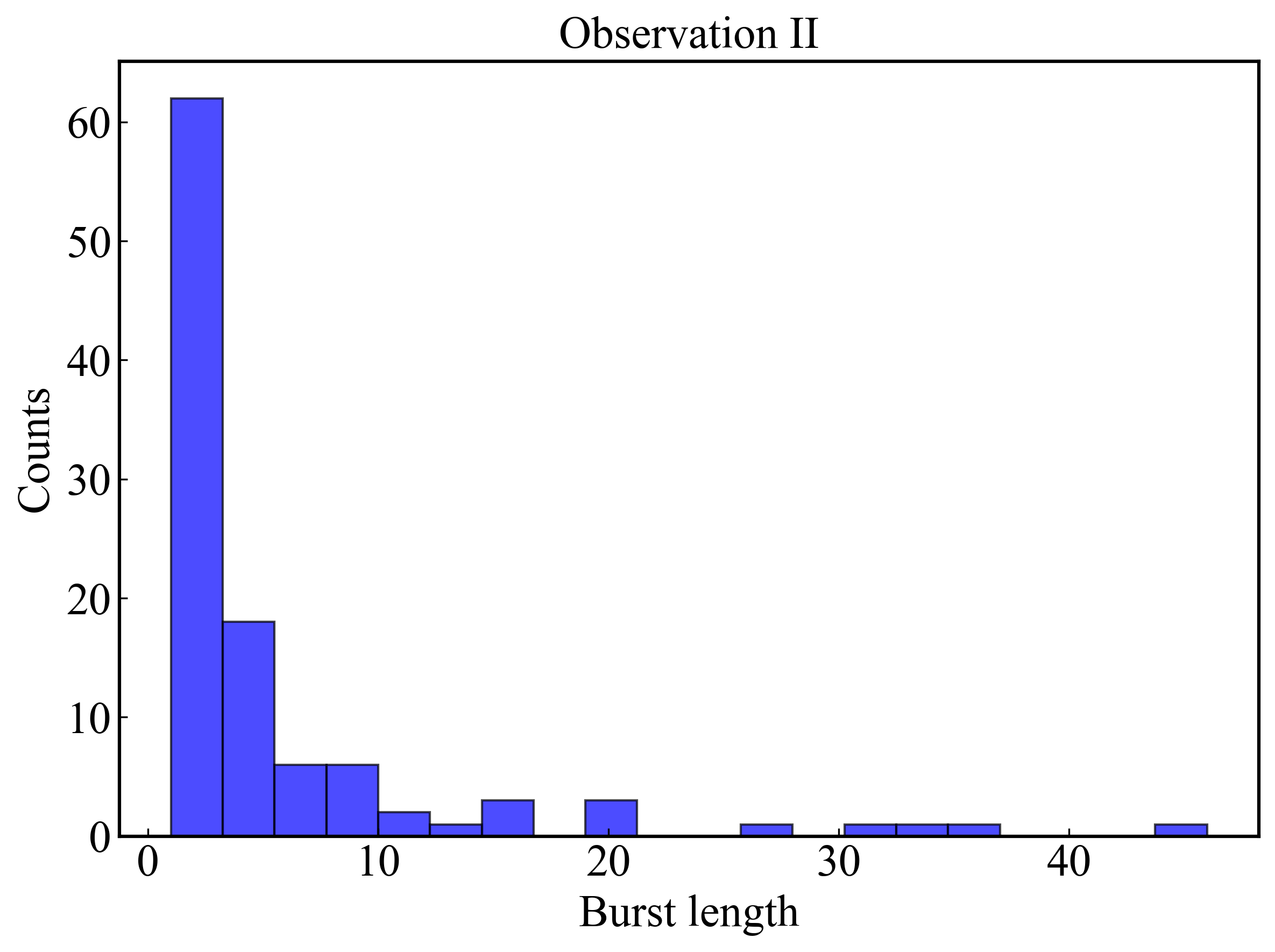}
        \caption{}
        \label{fig:burst II}
    \end{subfigure}\hfill
    \caption{Statistical distributions of burst lengths and nulling lengths for PSR~J1820$-$0509. The blue histograms (panels~c and~d) represent the burst-length distributions, while the red histograms (panels~a and~b) show the nulling-length distributions. The left panels (a,~c) correspond to the first observation, and the right panels (b,~d) correspond to the second observation. \textit{Null length:} In the second observation (panel~b), an extremely long nulling event spanning approximately 1750 pulses is detected, reflecting the complexity of radiation-state switching in this source. \textit{Burst length:} Both observations are dominated by short bursts. However, during the first observation, long bursts lasting more than 200 pulses are detected, whereas burst activities in the second observation are generally shorter ($<50$ pulses). The vertical axis is plotted on a logarithmic scale.}
    \label{fig:9}
\end{figure*}

To facilitate a clearer comparison between the two observations, we constructed histograms of null lengths and burst lengths following the approach described by \citet{cordes2013pulsar}.
Figure~\ref{fig:9} presents the distributions of burst lengths and nulling lengths that satisfy the selection criteria in the two data sets. The nulling-length distributions from the two observations are nearly identical overall, showing no substantial differences, except that the second observation exhibits a significantly larger number of counts near zero. This indicates that the number of pulse segments in the second observation is markedly higher than in the first, while the total number of pulses in the two observations is nearly the same. Consequently, the burst lengths in the second observation are significantly shorter than those in the first. This contrast is clearly reflected in the burst-length histograms: the first observation shows substantially longer burst lengths than \colorbold{black}{the second one}. In the first observation, long bursts are present with a maximum length of up to about 250 pulse periods, whereas in the second observation the maximum burst length reaches only around 50 periods. These results further demonstrate that PSR J1820$-$0509 exhibits two distinct states---dense burst clusters'' and sparse burst clusters''---across the two observations, and that its bursting activity shows pronounced temporal variability.

\begin{figure}
    \centering
    \begin{subfigure}{1.02\columnwidth}
        \centering
        \includegraphics[width=\columnwidth,height=0.6\textheight,keepaspectratio]{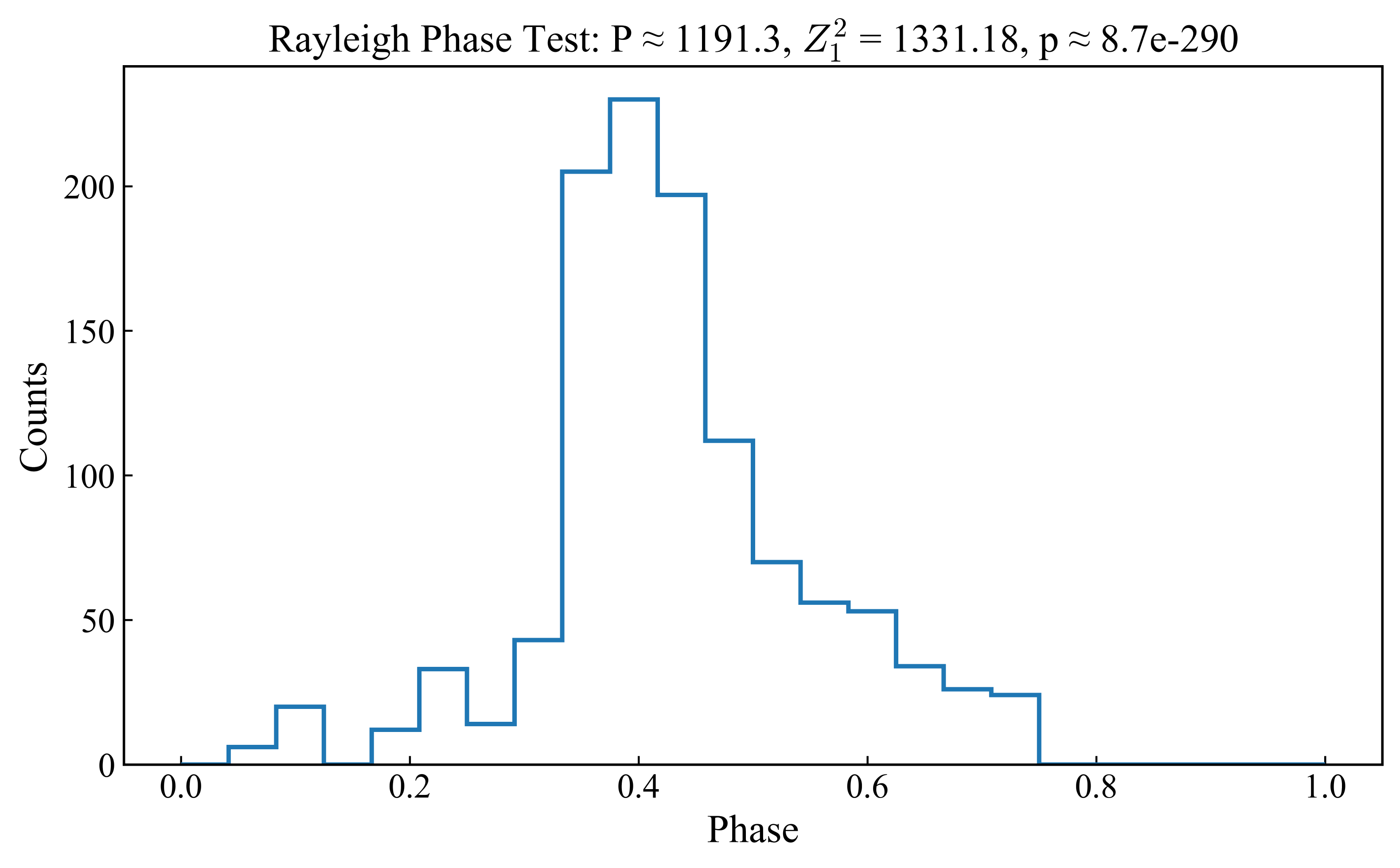}
        \caption{}
        \label{fig:Rayleigh phase test_a}
    \end{subfigure}
    \vskip\baselineskip
    \begin{subfigure}{1\columnwidth}
        \centering
        \includegraphics[width=\columnwidth,height=0.6\textheight,keepaspectratio]{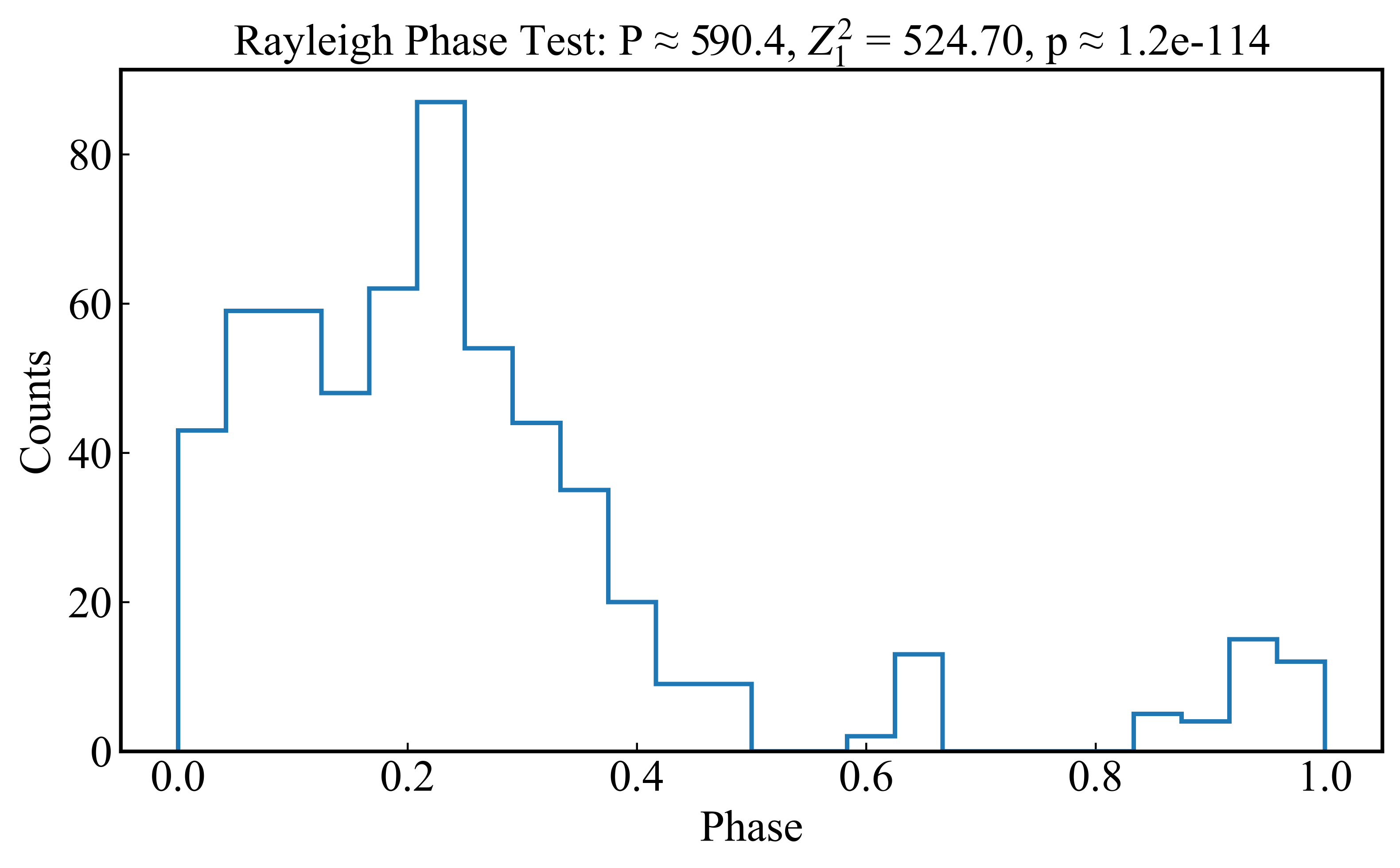}
        \caption{}
        \label{fig:Rayleigh phase test_b}
    \end{subfigure}

    \caption{The results of the Rayleigh test for phase clustering are summarized as follows. 
The blue solid line represents the histogram of the phase distribution. 
The title indicates the dominant periodicity $P$, the Rayleigh test statistic $Z_{1}^{2}$, 
and the corresponding $p$-value. 
Panel (a) shows the results from the first observation, with a dominant periodicity of 
$P \approx 1191.3 \pm 81$, $Z_{1}^{2} = 1331.18$, and $p \approx 8.7 \times 10^{-290}$. 
Panel (b) presents the results from the second observation, with a dominant periodicity of 
$P \approx 590.4 \pm 15$, $Z_{1}^{2} = 524.70$, and $p \approx 1.2 \times 10^{-114}$. 
Panel (a) corresponds to the first observation, while panel (b) corresponds to the second observation.
}
    \label{fig:Rayleigh phase test}
\end{figure}

To test the periodicity of the burst episode in PSR~J1820$-$0509, we simulated the random transition process between the ``burst state'' and ``null state'' using a Markov chain model \citep{cordes2013pulsar}, and constructed the empirical null distribution of the power spectrum. 
Based on this, we compared the power spectrum peak values of the two real observational sequences. Both observations displayed significant periodic peaks in their power spectra (Figure~\ref{fig:Rayleigh phase test_a},\ref{fig:Rayleigh phase test_b}).

In the first observation, the power spectrum reaches a peak (0.182395) at $P \approx 1179 \pm 81$ rotation periods , which significantly exceeds the 99\% global threshold of the null hypothesis distribution. Similarly, in the second observation, the power spectrum peak (0.062961) corresponds to a period of $P \approx 591 \pm 15$ rotation periods, also well above the corresponding 99\% threshold.The empirically estimated false-alarm probability (FAP) computed via Monte Carlo simulations is very low, suggesting that the observed peaks are unlikely to be the result of random fluctuations and providing strong evidence for the presence of genuine periodicity.

\begin{figure}
    \centering
    \begin{subfigure}{1\columnwidth}
        \centering
        \includegraphics[width=\columnwidth,height=0.6\textheight,keepaspectratio]{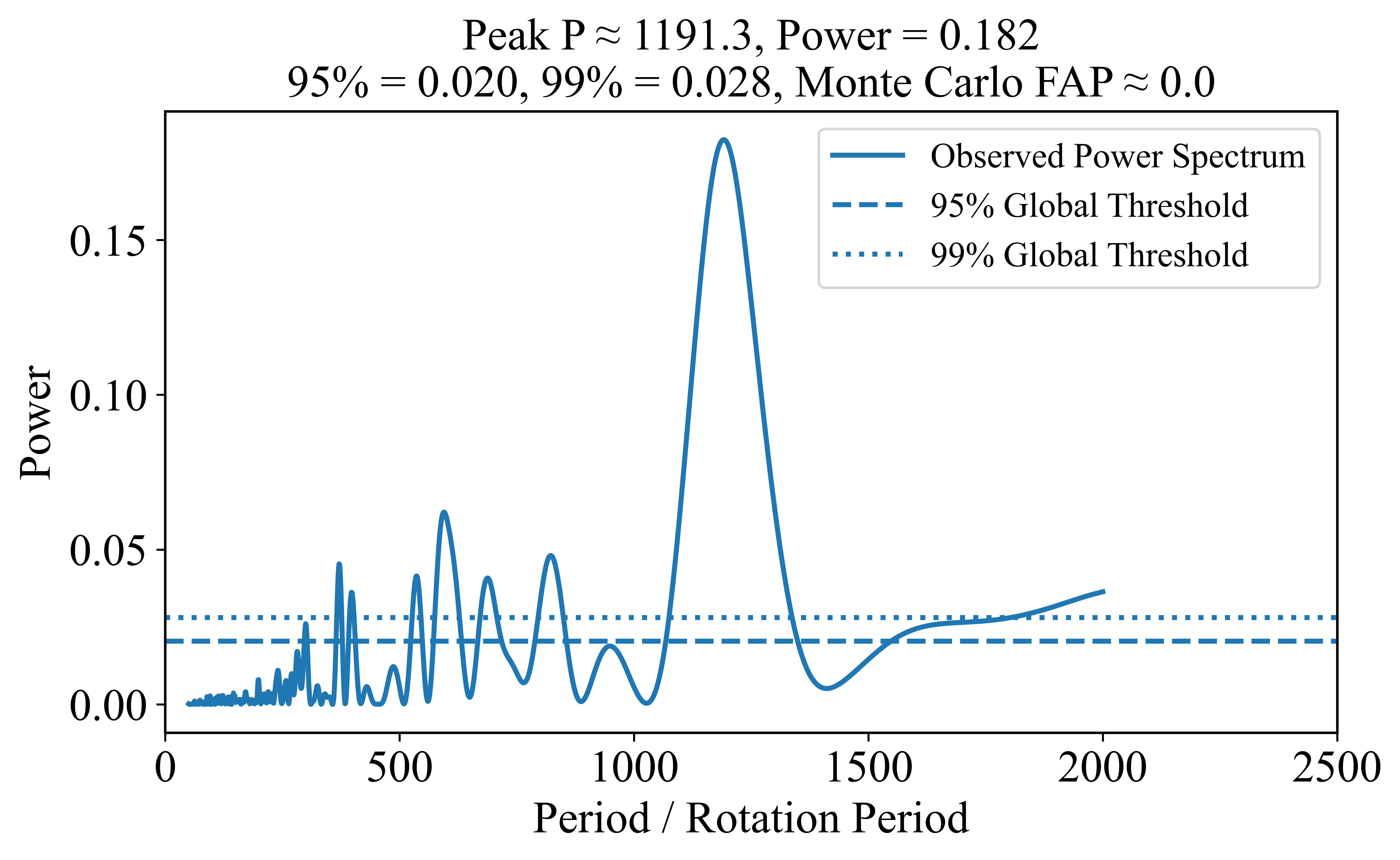}
        \caption{}
        \label{fig:Power spectrum_a}
    \end{subfigure}
    \vskip\baselineskip
    \begin{subfigure}{1\columnwidth}
        \centering
        \includegraphics[width=\columnwidth,height=0.6\textheight,keepaspectratio]{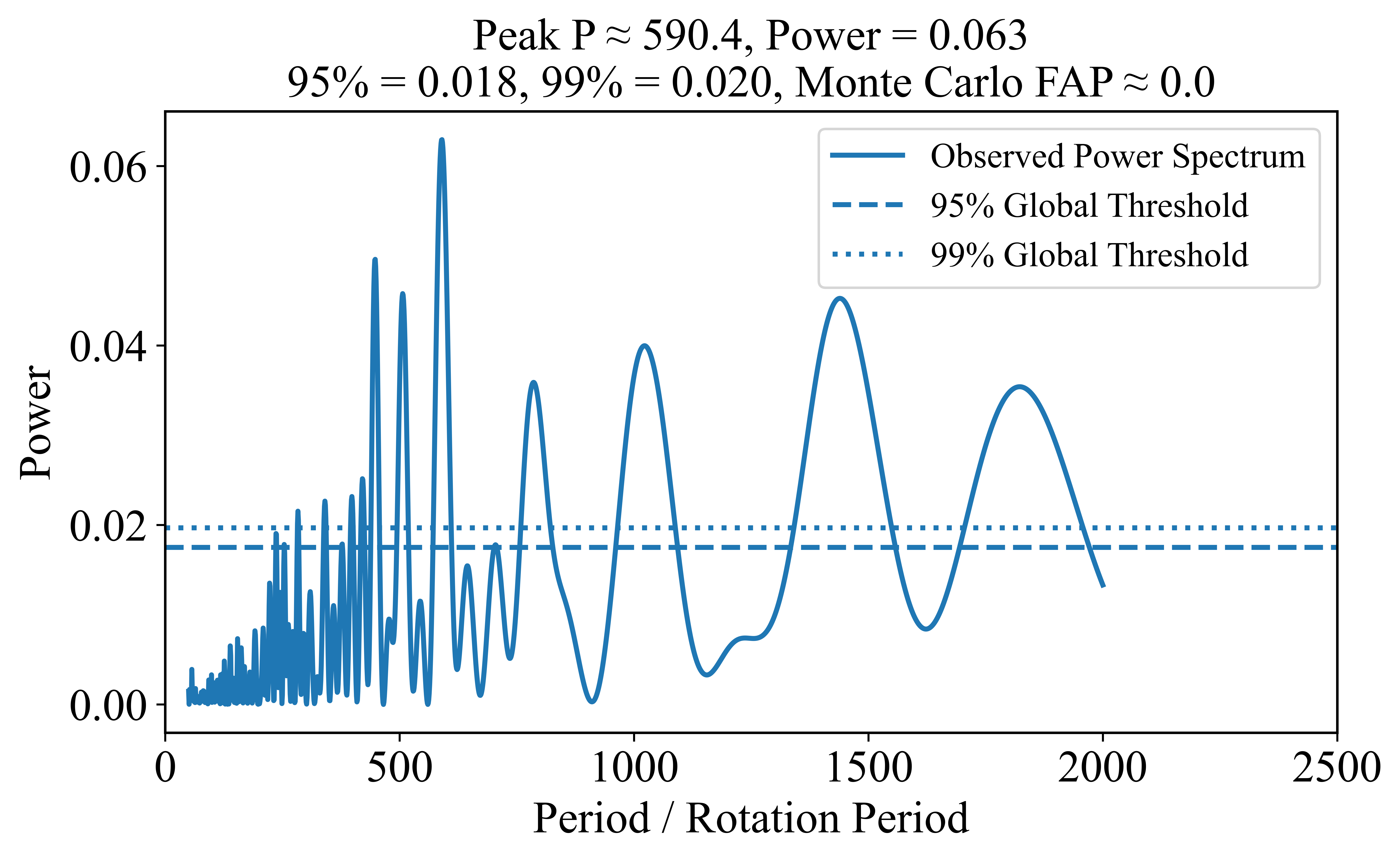}
        \caption{}
        \label{fig:Power spectrum_b}
    \end{subfigure}
    \caption{The power spectrum significance test for concentrated bursts is described as follows.The blue solid line represents the observed power spectrum, the blue dashed line marks the $99\%$ global threshold, and the yellow dashed line indicates the $95\%$ global threshold. Panel (a) shows the results from the first observation, with the main peak located at $P \approx 1191.3 \pm 81$ rotation periods, power of $0.182$, the $95\%$ global threshold of $0.020$, and the $99\%$ global threshold of $0.025$. Panel (b) presents the results from the second observation, with the main peak located at $P \approx 590.4 \pm 15$ rotation periods, power of $0.063$, the $95\%$ global threshold of $0.017$, and the $99\%$ global threshold of $0.018$.}
    \label{fig:Power spectrum}
\end{figure}

To further confirm the reliability of the candidate periods, we examine the phase distribution of burst events at the dominant period identified in the power spectrum. We test whether this phase distribution deviates significantly from a uniform distribution. For this purpose, we apply the first-order Rayleigh phase-concentration test, also known as the \( Z_1^2 \) statistic.\citep{de1989poweful,brazier1994confidence}.
First, a Lomb--Scargle power-spectrum analysis is performed on the pulse sequence, and the most significant peak is selected as the reference period $P_0$, expressed in units of the pulsar rotation period. The pulse index $n$ is then phase-folded at $P_0$, and the phase is defined as
\begin{equation}
\phi = \frac{n \bmod P_0}{P_0}, \qquad \phi \in [0,1).
\end{equation}
Only pulses identified as burst events are retained, and their corresponding phases are extracted to form the burst-phase set $\{\phi_i\}$. This yields the phase distribution of burst events within the unit phase interval (see Figure~\ref{fig:Power spectrum}).
To quantitatively assess whether this phase distribution departs from uniformity, we introduce the Rayleigh phase-concentration statistic, defined as
\begin{equation}
Z_1^2 = \frac{2}{N}
\left[
\left(\sum_{i=1}^{N} \cos 2\pi\phi_i \right)^2
+
\left(\sum_{i=1}^{N} \sin 2\pi\phi_i \right)^2
\right],
\end{equation}
where $N$ is the total number of burst pulses.
This statistic measures the squared length of the resultant vector formed by phase vectors on the unit circle and is therefore highly sensitive to phase clustering. Under the null hypothesis that burst events are uniformly distributed over the interval $[0,1)$, $Z_1^2$ follows a chi-square distribution with two degrees of freedom, and the corresponding significance probability is given by
\begin{equation}
p = \exp\left(-\frac{Z_1^2}{2}\right).
\end{equation}
A larger $Z_1^2$ value (or equivalently a smaller $p$-value) indicates significant phase concentration of burst events at the reference period $P_0$, whereas a smaller $Z_1^2$ implies that the phase distribution is consistent with uniformity.

Figure~\ref{fig:Rayleigh phase test} shows that, in the first observation, when folding the burst pulses at $P \approx 1191.3$ rotation periods, the test statistic is $Z_1^2 = 1331.18$ ($p \approx 8.65 \times 10^{-290}$), indicating a highly concentrated (non-random) phase distribution. Likewise, in the second observation, folding at $P \approx 590.4$ rotation periods yields $Z_1^2 = 524.70$ ($p \approx 1.16 \times 10^{-114}$), again confirming the strongly clustered nature of the phase distribution.

\begin{figure*}
    \centering
    \begin{minipage}{0.45\textwidth}
        \centering
        \includegraphics[width=\columnwidth,height=0.6\textheight,keepaspectratio]{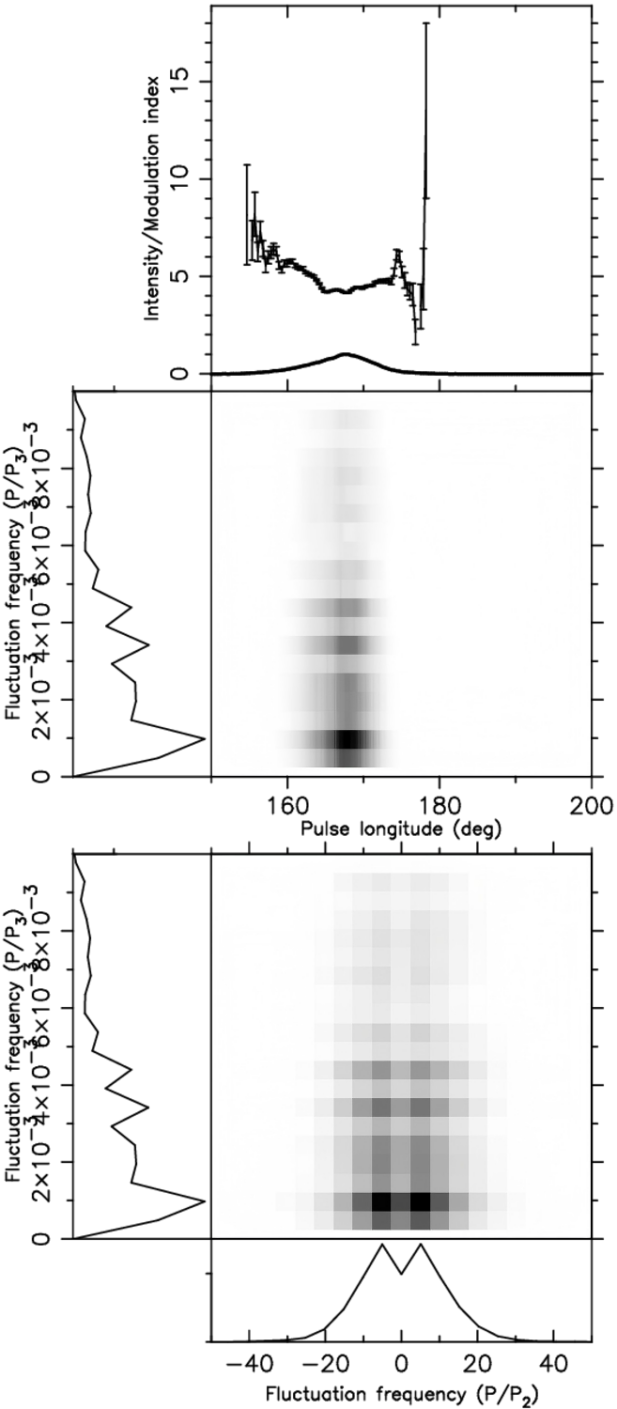}
        \\ Observation I
        \label{fig:2DFS_I}
    \end{minipage}\hfill
    \begin{minipage}{0.45\textwidth}
        \centering
        \includegraphics[width=\columnwidth,height=0.6\textheight,keepaspectratio]{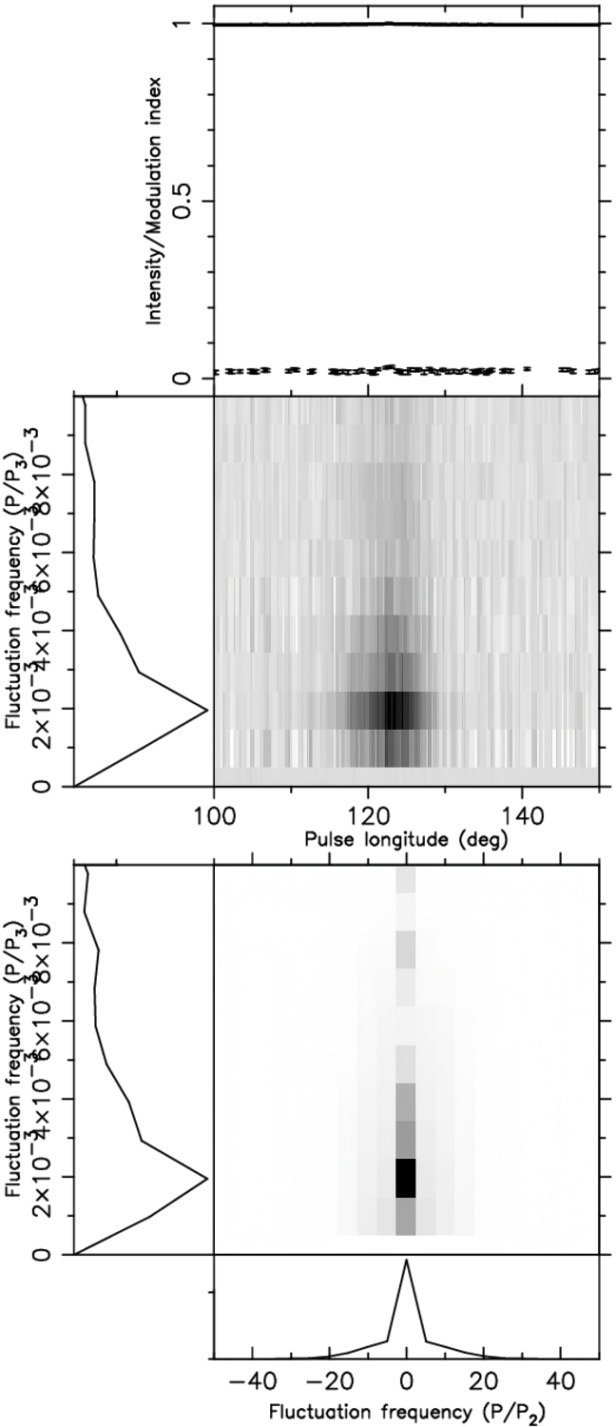}
        \\ Observation II
        \label{fig:2DFS_II}
    \end{minipage}\\[1ex]

    \caption{Longitude-resolved fluctuation spectra (LRFS) and two-dimensional fluctuation spectra (2DFS) obtained from two independent observations are shown. The left column corresponds to the first observation, while the right column corresponds to the second observation.In the first observation, a significant peak is detected in the 2DFS at a vertical fluctuation frequency of$\frac{P}{P_3} \approx 0.84 \times 10^{-3},$corresponding to a modulation period of $T \approx 1190\,P.$In contrast, during the second observation, this peak shifts to$\frac{P}{P_3} \approx 1.71 \times 10^{-3},$corresponding to a shorter modulation period of $ T \approx 585\,P.$ These results indicate that the pulsar exhibits markedly different subpulse modulation timescales at different observing epochs.}

    \label{fig:12}
\end{figure*}

To verify our periodicity results, we applied longitude-resolved and two-dimensional fluctuation spectra (LRFS and 2DFS; \citealt{weltevrede2016investigation}).
We performed fluctuation spectral analyses separately on the single-pulse sequences of the two emission states during the burst phases and computed the corresponding 2DFS (Figure~\ref{fig:12}). The results show that statistically significant periodic structures can be clearly identified in the spectra, indicating that the source indeed exhibits periodic modulation during the burst-emission phases.
We then compared the periodicities obtained from the LRFS/2DFS analysis with the statistical characteristics derived from the subsequent Markov-chain method. For the first observation, the peak at $\frac{P}{P_3} \approx 0.84 \times 10^{-3}$ corresponds to a period of  $T \approx 1190$,
while for the second observation, the peak at  $\frac{P}{P_3} \approx 1.71 \times 10^{-3}$ corresponds to a period of  $T \approx 585.$ The consistency between the two methods in characterizing emission-state transitions and periodic behavior provides independent support for the reliability of the statistical results obtained from the Markov-chain analysis.



\section{Discussion}

The nulling fraction of this pulsar reaches approximately 81.78\%, significantly higher than the $\sim$67\% reported in previous Parkes observations \citep{wang2007pulsar}. 
However, further analysis shows that these so-called ``null'' states still contain weak emission, indicating that radiation does not stop completely but remains below the detection threshold. This phenomenon is consistent with other extreme nulling pulsars (e.g., PSR~B0818$-$41; \citealt{bhattacharyya2010investigation}) and supports a continuous emission sequence comprising null, weak, and strong states.

The single-pulse analysis of PSR~J1820$-$0509 shows that its radio emission 
is regulated by multiple magnetospheric states and exhibits distinct 
statistical and polarization properties. The four emission states identified 
in the observations (Mode~A, Mode~B, Mode~C, and Mode~D (a pseudo-nulling state)) 
indicate that the magnetosphere of this pulsar can switch between several 
quasi-stable configurations, which likely correspond to different particle 
flow conditions and radiation coherence properties (e.g., 
\citet{lyne2010switched,kramer2006periodically}).

The statistical relation between the single-pulse $3\sigma$ width and the 
relative energy provides further evidence for the presence of different 
emission modulation mechanisms. The width--relative energy relation shows 
a transition around a characteristic energy scale, indicating that once 
the pulse relative energy exceeds this threshold, the variation pattern 
of the pulse width changes systematically. This behaviour may reflect a 
change in the effective size of the emission region, or that the beam 
structure tends to approach saturation at higher energy states.
In addition, the energy distributions of different emission modes exhibit 
distinct statistical forms. The energy distributions of Mode~A and Mode~C 
follow lognormal distributions, which are commonly associated with 
multiplicative stochastic processes and have been widely reported in 
single-pulse studies of many pulsars 
(\citealt{cairns2001intrinsic,burke2012high}). By contrast, the energy distribution of Mode~B shows a composite structure consisting of both Gaussian and lognormal components, suggesting that this state may involve the coexistence of a relatively 
stable emission component and a stochastic fluctuation-driven component. 
It is noteworthy that although the occurrence fractions of the different 
emission modes vary between the two observations, the statistical 
properties of the energy distributions corresponding to each mode remain 
highly consistent between the two epochs. This indicates that the 
underlying emission mechanisms themselves are stable, while the overall 
variability of the pulsar emission is mainly caused by changes in the 
residence time of the magnetosphere in different states.
The energy distribution of Mode~D (the pseudo-nulling state) is close to a Gaussian distribution centered near zero, indicating that the observed signal in this state is primarily dominated by system noise.

The polarization properties differ significantly among the emission modes. In Mode~A, the emission is generally weak, with a highly scattered PPA distribution and a low degree of linear polarization, exhibiting a clear depolarization feature. This behavior is commonly interpreted as the result of the superposition of OPMs (\citealt{stinebring1984pulsara,stinebring1984pulsarb,mckinnon2000mode}) and may also reflect enhanced magnetospheric plasma propagation effects or local perturbations. In contrast, Mode~B shows the most ordered polarization structure: the circular polarization exhibits a clear sign reversal near the main pulse peak, while the PPA varies smoothly with pulse phase. This behavior is broadly consistent with the RVM (\citealt{radhakrishnan1969evidence}), indicating that the emission geometry is relatively stable in this state. Mode~C, on the other hand, appears as sporadic isolated burst events. These pulses display a pronounced negative circular polarization extremum near the main peak, together with a relatively low fraction of linear polarization. Similar polarization behavior has been reported in single-pulse studies of several pulsars and is often attributed to magnetospheric propagation effects or changes in the geometry of the emission region \citep{karastergiou2003simultaneous}.

Another remarkable property of PSR~J1820$-$0509 is the presence of clustered 
burst activities lasting from several tens to a few hundred rotation periods. 
The duration of the burst episodes differs significantly between the two 
observations. In the first observation, the longest burst episode extends to 
about 250 rotation periods, whereas in the second observation the longest one 
lasts only about 50 periods, indicating clear temporal variability.
Nevertheless, both data sets reveal quasi-periodic modulation within the burst 
activities. Power spectral analysis identifies significant periodicities of 
approximately $(1191 \pm 81)$ and $(590 \pm 15)$ rotation periods in the two 
observing epochs, respectively. The significance of these periodic peaks is 
confirmed through Markov-chain modelling, Monte Carlo simulations, the 
Rayleigh phase test, and LRFS/2DFS analyses, demonstrating that the observed 
modulation is unlikely to arise from random fluctuations.
These periods are much longer than typical subpulse drifting periods 
(usually a few to several tens of rotation periods), and therefore are 
unlikely to be associated with the classical carousel-type subpulse drifting 
mechanism(\citealt{zhao2023single}). Similar periodic or quasi-periodic intensity modulations have 
also been reported in other pulsars. For example, some sources exhibit 
periodic amplitude modulation without longitude-dependent drifting 
(\citealt{mitra2017periodic}), while statistical studies of single pulses have shown 
that certain pulsars display long-timescale intensity oscillations unrelated 
to subpulse drifting (\citealt{basu2020periodic,basu2016meterwavelength}). In high-sensitivity 
observations with FAST, periodic amplitude modulation has also been reported 
to be associated with emission mode changes (\citealt{yan2020periodic}).
Compared with these phenomena, the periodic behaviour observed in 
PSR~J1820$-$0509 appears to be more consistent with an organized transition 
of the global emission state between burst and pseudo-nulling states, rather 
than a longitude-resolved drifting structure. Our LRFS and 2DFS analyses do 
not reveal clear systematic drifting tracks, but instead show an enhancement 
of low-frequency modulation components, further supporting the interpretation 
that the observed behaviour is dominated by a global intensity modulation 
process.

In terms of physical interpretation, the pulsar magnetosphere may possess 
multiple quasi-stable current-distribution states and can undergo transitions 
between them, thereby producing nulling or mode changing 
(\citealt{timokhin2010model,lyne2010switched}). Within the framework of the partially 
screened gap (PSG) model, the thermal--electric feedback process in the 
polar-cap region may also modify the discharge conditions and lead to 
changes in the emission state (\citealt{geppert2021rapid}). These theoretical 
considerations provide a possible physical background for understanding 
the long-timescale modulation of emission states.
Therefore, the quasi-periodic burst activities observed in PSR~J1820$-$0509 
are likely to reflect periodic reconfigurations of the magnetospheric 
structure or particle discharge conditions on macroscopic timescales. 
When the magnetosphere is in a stable discharge configuration, the pulsar 
manifests strong burst emission. In contrast, when the plasma supply is 
reduced or the magnetospheric structure becomes unstable, the emission 
intensity decreases and the pulsar enters a weak-emission or 
pseudo-nulling state.
Similar quasi-periodic burst behaviours have also been reported in other 
pulsars. For instance, PSR~B1931+24 exhibits periodic transitions between 
emission and quiescent phases, with activity cycles lasting several days 
(\citealt{kramer2006periodically}), while PSR~J1107$-$5907 alternates between strong 
emission episodes and long-duration weak-emission phases 
(\citealt{young2014apparent}). These sources indicate that pulsar emission 
variability may be closely related to changes in the magnetospheric plasma 
dynamics.
Overall, PSR~J1820$-$0509 does not simply switch between an emission state 
and a null state, but instead exhibits multi-mode emission behaviour 
modulated by quasi-periodic magnetospheric activity. These observations 
provide new empirical constraints on the emission mechanisms of high-nulling 
pulsars and further highlight the importance of high-sensitivity 
single-pulse observations for probing plasma processes in pulsar 
magnetospheres.

\begin{figure}
    \centering
    \begin{subfigure}{1\columnwidth}
        \includegraphics[width=\columnwidth,height=0.6\textheight,keepaspectratio]{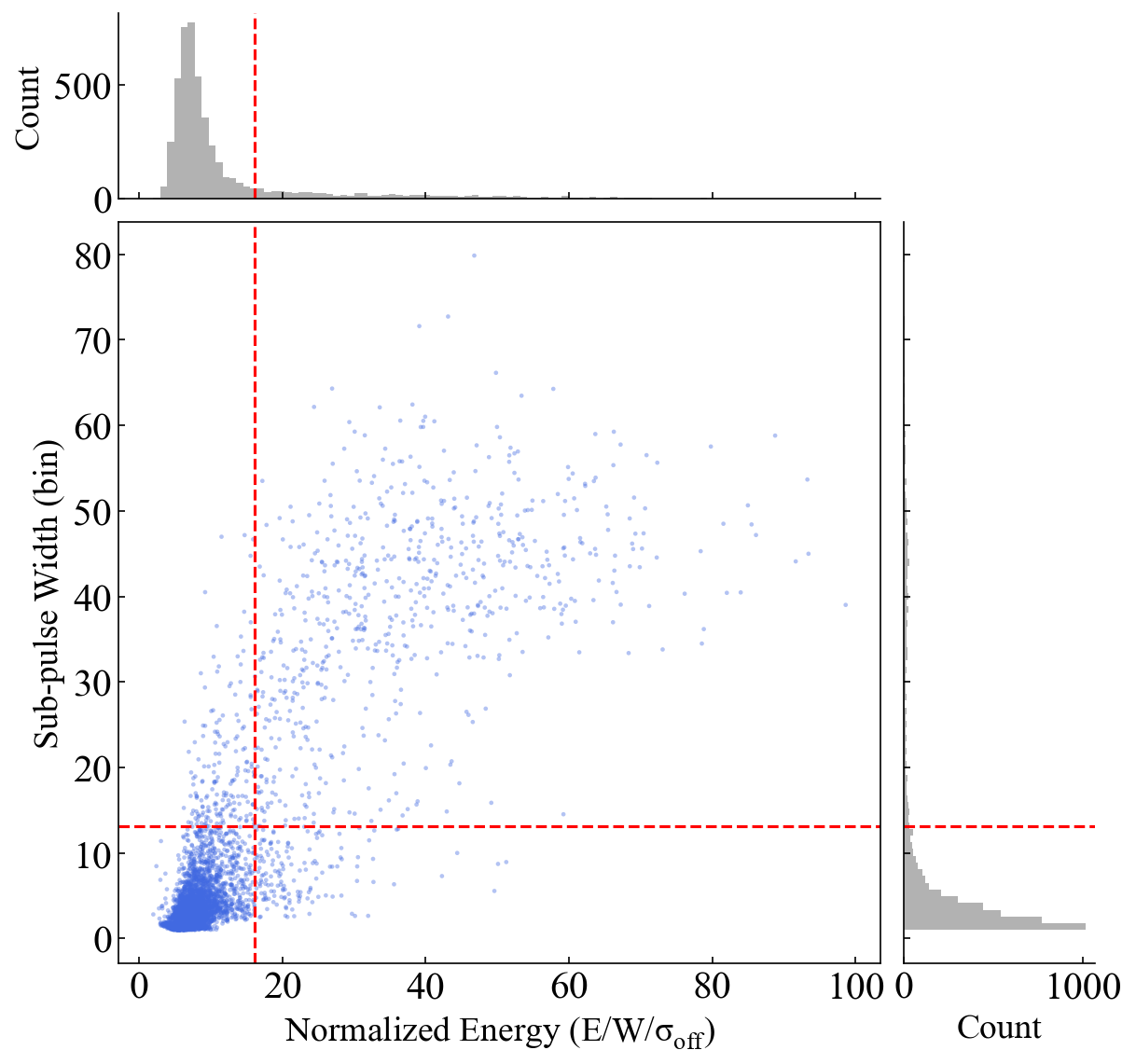}
        \label{fig:draw_pulsar}
    \end{subfigure}
    \hfill
    \caption{The central scatter plot shows the properties of the detected single-pulse events, with the horizontal axis representing the normalized energy, defined as $E_{\rm norm} = \frac{E/W}{\sigma_{\rm off}},$ where the mean intensity of each pulse within the on-pulse window is divided by the rms $\sigma_{\rm off}$ calculated from the off-pulse region. The vertical axis represents the subpulse width in units of bins. The top and right panels show the marginal distributions of energy and width, respectively. The red dashed line indicates the $3\sigma$ threshold used to select dwarf-pulse candidates. A dense cluster of points is clearly visible in the lower-left corner (low energy and narrow width), corresponding to the dwarf-pulse population defined by \citep{yan2024dwarf}. These pulses exhibit extremely low emission energy and very narrow widths, which are clearly separated from the higher-energy normal pulses or background noise distribution.}
    \label{fig:15}
\end{figure}

High-sensitivity observations with FAST have revealed, in several pulsars, a form of emission that lies between normal pulses and complete radio silence, commonly referred to as dwarf pulses \citep{chen2023strong,yan2024dwarf}. Such emission typically occurs during intervals traditionally classified as null states. Although the single-pulse energies are significantly lower than those of the normal emission mode, their temporal structures and pulse-phase locations remain associated with the main pulse window.
In this work, we likewise detect a persistent low-level emission component during the null state of PSR~J1820$-$0509. Figure~\ref{fig:15} shows the joint distribution of normalized single-pulse energy and subpulse width \citep{yan2024dwarf}. It is evident that this low-energy emission is not simply mixed with background noise, but instead forms a statistically distinguishable population in parameter space. As shown in the figure, a substantial number of detectable emission events remain in the low normalized-energy regime (to the left of the red vertical dashed line), with an energy distribution that clearly deviates from pure Gaussian noise. This behavior closely resembles the energy characteristics of dwarf pulses reported by \cite{yan2024dwarf}, in which dwarf pulses do not form a distinct high-energy tail separated from normal pulses, but instead appear as a low-energy, numerically dominant population. From a statistical perspective, the low-level emission observed during the null state of this source therefore shares the basic energy properties characteristic of known dwarf-pulse phenomena.
The distribution of subpulse widths provides further insight into the nature of the emission. As shown in Figure~\ref{fig:15}, the low-energy emission is predominantly associated with relatively narrow subpulse widths, while the subpulse width increases systematically with increasing energy. This ``low-energy--narrow-width'' correspondence is consistent with the characterization of dwarf pulses given by \cite{chen2023strong}, where dwarf pulses are typically short-lived and structurally simple events, in contrast to the broader and more complex substructures observed in the normal emission state. Notably, such a width--energy correlation is inconsistent with expectations from random noise or baseline fluctuations, further supporting a genuine astrophysical origin for the low-level emission.
Taken together with previous studies and the present results, it is plausible that a fraction of the events classified here as ``low-level emission'' would be re-identified as discrete dwarf pulses under observations with higher sensitivity or finer time resolution. This interpretation is consistent with FAST observations of other pulsars, which have shown that the null state is not completely radio-quiet, but instead decomposes into a statistical ensemble of numerous low-energy, narrow-width emission events as sensitivity increases. However, we emphasize that we do not perform a direct, event-by-event morphological identification of dwarf pulses in this work. Accordingly, we do not explicitly classify these events as  dwarf  pulses, but rather note their strong statistical similarity to previously reported dwarf-pulse phenomena.
With the current data, it is not yet possible to determine at the level of physical emission mechanisms whether this low-level emission arises from the same process responsible for the dwarf pulses described in the literature. The present results instead favor an interpretation in which the null state does not correspond to a complete shutdown of the radio emission mechanism, but rather to a magnetospheric state in which particle acceleration or coherent emission efficiency is substantially reduced, though not entirely suppressed. Whether an additional low-level emission mechanism exists that is distinct from dwarf pulses will require higher signal-to-noise single-pulse morphological analyses and systematic comparisons of polarization properties.

\section{Conclusion}

In this study, we performed two high-sensitivity single-pulse observations 
(1.05--1.45~GHz) of PSR~J1820$-$0509 using the Five-hundred-meter Aperture 
Spherical radio Telescope (FAST), analyzing over 17,000 pulses to investigate 
its nulling behaviour, energy statistics,emission mode 
changes, and polarization properties, burst activities. The measured nulling fraction is approximately 81.78\%, significantly higher than previous Parkes observations. 
Stacking analyses of pulses classified as nulls reveal a signal above the 
baseline level outside the pulse phase, indicating the presence of persistent 
weak emission during the null intervals, consistent with pseudo-nulling 
behaviour observed in other high-nulling pulsars.

Four distinct emission modes were identified in the single-pulse sequences: 
Mode~A-relatively weak, multi-component emission; 
Mode~B-stronger, single-component emission; 
Mode~C-short-duration, isolated burst events; 
and Mode~D-single pulses with emission energies below the detection threshold, 
while the stacked average profile still exhibits a significant signal-to-noise 
ratio. Modes~A, B, and C exhibit clear differences in pulse morphology, 
polarization properties, and energy statistics, indicating that they correspond 
to different quasi-stable magnetospheric states. A statistically significant 
transition is observed in the pulse width versus normalized pulse energy 
relationship. Piecewise linear regression reveals two distinct regions 
separated by a characteristic transition energy in each emission mode. The 
energy distributions of the three modes follow distinct statistical forms: 
Modes~A and C are well described by lognormal distributions, whereas Mode~B 
displays a composite distribution composed of Gaussian and lognormal components.

Long pulse sequences reveal pronounced burst activities lasting tens to 
hundreds of rotation periods. Through Markov chain simulations, power 
spectrum analysis, and Rayleigh phase concentration tests, quasi-periodic 
modulations are identified, with characteristic periods of $1191 \pm 81$ and 
$590 \pm 15$ rotation periods in the two observations. 

Overall, PSR~J1820$-$0509 exhibits a continuum of emission from pseudo-nulling 
and weak-emission states to strong burst-dominated states. The observed 
energy distributions, burst modulations, mode transitions, and polarization 
behaviours provide new constraints on the dynamics and nonlinear characteristics 
of pulsar magnetospheric emission, particularly for pulsars with extremely 
high nulling fractions.

\section*{Acknowledgments}
We hereby acknowledge support from the following agencies and programs: National Natural Science Foundation of China (Grant Nos. 12563008, 11988101, U1731238, U2031117, 11565010, 11725313, 1227308, 12041303, 12588202); National Key R\&D Program of China (No. 2023YFE0110500); National SKA Program of China (Nos. 2020SKA0120200, 2022SKA0130100, 2022SKA0130104); Science and Technology Foundation of Guizhou Provincial Department of Education (No. KY(2023)059); Youth Innovation Promotion Association CAS (ID: 2021055); Youth Scientists Project of Basic Research in CAS (YSBR-006); Foreign Talents Program (No. QN2023061004L; E.G.); and the Scientific and Technological Innovation Team of Higher Education Institutions under the Education Department of Guizhou of China (Grant No. QJJ[2023]093); CAS Youth Interdisciplinary Team; Liupanshui Science and Technology Development Project (No. 52020-2024-PT-01); and the Cultivation Project for FAST Scientific Payoff and Research Achievement of CAMS-CAS. P.W. acknowledges additional support from the CAS Youth Interdisciplinary Team, Youth Innovation Promotion Association CAS, and the Cultivation Project for FAST Scientific Payoff and Research Achievement of CAMS-CAS. D.L. is supported as a New Cornerstone Investigator. These supports were instrumental in the successful completion of this work.

\section*{Data availability}
The data underlying this article will be shared on reasonable request to the corresponding author.


\bibliographystyle{mnras}
\bibliography{example} 

@article{hewish1968pulsars,
  title={Pulsars},
  author={Hewish, Antony},
  journal={Scientific American},
  volume={219},
  number={4},
  pages={25--35},
  year={1968},
  publisher={JSTOR}
}

@article{chen2023meerkat,
  title={MeerKAT discovery of 13 new pulsars in Omega Centauri},
  author={Chen, Weiwei and Freire, PCC and Ridolfi, A and Barr, ED and Stappers, B and Kramer, M and Possenti, A and Ransom, SM and Levin, L and Breton, RP and others},
  journal={Monthly Notices of the Royal Astronomical Society},
  volume={520},
  number={3},
  pages={3847--3856},
  year={2023},
  publisher={Oxford University Press}
}

@article{liu2015single,
  title={Single-pulse and profile-variability study of PSR J1022+ 1001},
  author={Liu, Kuo and Karuppusamy, R and Lee, KJ and Stappers, BW and Kramer, M and Smits, R and Purver, MB and Janssen, GH and Perrodin, DELPHINE},
  journal={Monthly Notices of the Royal Astronomical Society},
  volume={449},
  number={1},
  pages={1158--1169},
  year={2015},
  publisher={Oxford University Press}
}

@article{palliyaguru2023single,
  title={Single-pulse studies of three millisecond pulsars},
  author={Palliyaguru, NT and Perera, BBP and McLaughlin, MA and Os{\l}owski, S and Siebert, GL},
  journal={Monthly Notices of the Royal Astronomical Society},
  volume={520},
  number={2},
  pages={2747--2756},
  year={2023},
  publisher={Oxford University Press}
}

@article{uttley2005non,
  title={Non-linear X-ray variability in X-ray binaries and active galaxies},
  author={Uttley, Phil and McHardy, IM and Vaughan, S},
  journal={Monthly Notices of the Royal Astronomical Society},
  volume={359},
  number={1},
  pages={345--362},
  year={2005},
  publisher={Blackwell Science Ltd Oxford, UK}
}

@article{stinebring1984pulsara,
  title={Pulsar polarization fluctuations. I-1404 MHz statistical summaries},
  author={Stinebring, Daniel R and Cordes, JM and Rankin, Joanna M and Weisberg, JM and Boriakoff, Valentin},
  journal={Astrophysical Journal Supplement Series (ISSN 0067-0049), vol. 55, June 1984, p. 247-277},
  volume={55},
  pages={247--277},
  year={1984}
}

@article{stinebring1984pulsarb,
  title={Pulsar Polarization Fluctuations. II-800-MHZ Statistical Summaries},
  author={Stinebring, DR and Cordes, JM and Weisberg, JM and Rankin, JM and Boriakoff, V},
  journal={Astrophysical Journal Supplement, Vol. 55, Issue 2, p. 279-288, 1984},
  volume={55},
  pages={279--288},
  year={1984}
}

@article{rankin1990toward,
  title={Toward an empirical theory of pulsar emission. IV-Geometry of the core emission region},
  author={Rankin, Joanna M},
  journal={Astrophysical Journal, Part 1 (ISSN 0004-637X), vol. 352, March 20, 1990, p. 247-257. Research supported by NSF, Research Corp., and University of Vermont.},
  volume={352},
  pages={247--257},
  year={1990}
}

@article{wang2010polarization,
  title={Polarization changes of pulsars due to wave propagation through magnetospheres},
  author={Wang, Chen and Lai, Dong and Han, JinLin},
  journal={Monthly Notices of the Royal Astronomical Society},
  volume={403},
  number={2},
  pages={569--588},
  year={2010},
  publisher={Blackwell Publishing Ltd Oxford, UK}
}

@article{mitra2023evidence,
  title={Evidence for coherent curvature radiation in PSR J1645- 0317 with disordered distribution of polarization position angle},
  author={Mitra, Dipanjan and Melikidze, George I and Basu, Rahul},
  journal={Monthly Notices of the Royal Astronomical Society: Letters},
  volume={521},
  number={1},
  pages={L34--L38},
  year={2023},
  publisher={Oxford University Press}
}

@article{manchester1975observations,
  title={Observations of pulsar radio emission. II-Polarization of individual pulses},
  author={Manchester, RN and Taylor, JH and Huguenin, GR},
  journal={Astrophysical Journal, vol. 196, Feb. 15, 1975, pt. 1, p. 83-102.},
  volume={196},
  pages={83--102},
  year={1975}
}

@article{smits2006frequency,
  title={Frequency dependence of orthogonal polarisation modes in pulsars},
  author={Smits, JM and Stappers, BW and Edwards, RT and Kuijpers, J and Ramachandran, R},
  journal={Astronomy \& Astrophysics},
  volume={448},
  number={3},
  pages={1139--1148},
  year={2006},
  publisher={EDP Sciences}
}

@article{lyne2010switched,
  title={Switched magnetospheric regulation of pulsar spin-down},
  author={Lyne, Andrew and Hobbs, George and Kramer, Michael and Stairs, Ingrid and Stappers, Ben},
  journal={Science},
  volume={329},
  number={5990},
  pages={408--412},
  year={2010},
  publisher={American Association for the Advancement of Science}
}

@article{radhakrishnan1969evidence,
  title={Evidence in support of a rotational model for the pulsar PSR 0833--45},
  author={Radhakrishnan, V and Cooke, DJ and Komesaroff, MM and Morris, D},
  journal={Nature},
  volume={221},
  number={5179},
  pages={443--446},
  year={1969},
  publisher={Nature Publishing Group UK London}
}

@article{mckinnon2000mode,
  title={The mode-separated pulse profiles of pulsar radio emission},
  author={McKinnon, Mark M and Stinebring, Daniel R},
  journal={The Astrophysical Journal},
  volume={529},
  number={1},
  pages={435--446},
  year={2000}
}

@article{mitra2017periodic,
  title={Periodic longitude-stationary non-drift emission in core-single radio pulsar B1946+ 35},
  author={Mitra, Dipanjan and Rankin, Joanna},
  journal={Monthly Notices of the Royal Astronomical Society},
  volume={468},
  number={4},
  pages={4601--4609},
  year={2017},
  publisher={Oxford University Press}
}

@article{yan2020periodic,
  title={Periodic mode changing in PSR J1048- 5832},
  author={Yan, WM and Manchester, RN and Wang, N and Wen, ZG and Yuan, JP and Lee, KJ and Chen, JL},
  journal={Monthly Notices of the Royal Astronomical Society},
  volume={491},
  number={4},
  pages={4634--4641},
  year={2020},
  publisher={Oxford University Press}
}

@article{timokhin2010model,
  title={A model for nulling and mode changing in pulsars},
  author={Timokhin, AN},
  journal={Monthly Notices of the Royal Astronomical Society: Letters},
  volume={408},
  number={1},
  pages={L41--L45},
  year={2010},
  publisher={Blackwell Science Ltd Oxford, UK}
}

@article{geppert2021rapid,
  title={Rapid modification of neutron star surface magnetic field: a proposed mechanism for explaining radio emission state changes in pulsars},
  author={Geppert, U and Basu, R and Mitra, D and Melikidze, GI and Szkudlarek, M},
  journal={Monthly Notices of the Royal Astronomical Society},
  volume={504},
  number={4},
  pages={5741--5753},
  year={2021},
  publisher={Oxford University Press}
}

@article{zhao2023single,
  title={Single-pulse behaviours and fast radio burst-like micropulses in FAST wide-band observations of eight pulsars},
  author={Zhao, Rushuang and Li, Di and Hobbs, George and Wang, Pei and Xue, Mengyao and Dang, Shijun and Liu, Hui and Zhi, Qijun and Miao, Chenchen and Yuan, Mao and others},
  journal={Monthly Notices of the Royal Astronomical Society},
  volume={521},
  number={2},
  pages={2298--2325},
  year={2023},
  publisher={Oxford University Press}
}

@article{cordes2013pulsar,
  title={Pulsar state switching from Markov transitions and stochastic resonance},
  author={Cordes, JM},
  journal={The Astrophysical Journal},
  volume={775},
  number={1},
  pages={47},
  year={2013},
  publisher={IOP Publishing}
}

@article{radhakrishnan1969magnetic,
author  = {Radhakrishnan, V. and Cooke, D.~J.},
  title   = {Magnetic poles and the polarization structure of pulsar radiation},
  journal = {Astrophysical Letters},
  year    = {1969},
  volume  = {3},
  pages   = {225--229},
  bibcode = {1969ApL.....3..225R}
}

@article{de1989poweful,
  title={A poweful test for weak periodic signals with unknown light curve shape in sparse data},
  author={De Jager, OC and Raubenheimer, BC and Swanepoel, JWH},
  journal={Astronomy and Astrophysics (ISSN 0004-6361), vol. 221, no. 1, Aug. 1989, p. 180-190.},
  volume={221},
  pages={180--190},
  year={1989}
}

@article{brazier1994confidence,
  title={Confidence intervals from the Rayleigh test},
  author={Brazier, KTS},
  journal={Monthly Notices of the Royal Astronomical Society},
  volume={268},
  number={3},
  pages={709--712},
  year={1994},
  publisher={Oxford University Press Oxford, UK}
}

@article{weltevrede2016investigation,
  title={Investigation of the bi-drifting subpulses of radio pulsar B1839- 04 utilising the open-source data-analysis project PSRSALSA},
  author={Weltevrede, Patrick},
  journal={Astronomy \& Astrophysics},
  volume={590},
  pages={A109},
  year={2016},
  publisher={EDP Sciences}
}

@article{rahaman2021mode,
  title={Mode changing, subpulse drifting, and nulling in four component conal pulsar PSR J2321+ 6024},
  author={Rahaman, S k Minhajur and Basu, Rahul and Mitra, Dipanjan and Melikidze, George I},
  journal={Monthly Notices of the Royal Astronomical Society},
  volume={500},
  number={3},
  pages={4139--4152},
  year={2021},
  publisher={Oxford University Press}
}

@article{basu2023single,
  title={Single pulse emission from PSR B0809+ 74 at 150 MHz using Polish LOFAR station},
  author={Basu, Rahul and Lewandowski, Wojciech and Kijak, Jaros{\l}aw and Bartosz, {\'S}mierciak and Soida, Marian and B{\l}aszkiewicz, Leszek and Krankowski, Andrzej},
  journal={Monthly Notices of the Royal Astronomical Society},
  volume={526},
  number={1},
  pages={691--699},
  year={2023},
  publisher={Oxford University Press}
}

@article{bhattacharyya2010investigation,
  title={Investigation of the unique nulling properties of PSR B0818- 41},
  author={Bhattacharyya, Bhaswati and Gupta, Yashwant and Gil, Janusz},
  journal={Monthly Notices of the Royal Astronomical Society},
  volume={408},
  number={1},
  pages={407--421},
  year={2010},
  publisher={Blackwell Publishing Ltd Oxford, UK}
}

@article{cairns2001intrinsic,
  title={Intrinsic variability of the Vela pulsar: Lognormal statistics and theoretical implications},
  author={Cairns, Iver H and Johnston, S and Das, P},
  journal={The Astrophysical Journal},
  volume={563},
  number={1},
  pages={L65},
  year={2001},
  publisher={IOP Publishing}
}

@article{burke2012high,
  title={The high time resolution universe pulsar survey--v. single-pulse energetics and modulation properties of 315 pulsars},
  author={Burke-Spolaor, S and Johnston, S and Bailes, Matthew and Bates, SD and Bhat, NDR and Burgay, M and Champion, DJ and D’Amico, Nicolo' and Keith, MJ and Kramer, M and others},
  journal={Monthly Notices of the Royal Astronomical Society},
  volume={423},
  number={2},
  pages={1351--1367},
  year={2012},
  publisher={Blackwell Publishing Ltd Oxford, UK}
}

@article{karastergiou2003simultaneous,
  title={Simultaneous single-pulse observations of radio pulsars-III. The behaviour of circular polarization},
  author={Karastergiou, Aris and Johnston, S and Kramer, M},
  journal={Astronomy \& Astrophysics},
  volume={404},
  number={1},
  pages={325--332},
  year={2003},
  publisher={EDP Sciences}
}

@article{chen2023strong,
  title={Strong and weak pulsar radio emission due to thunderstorms and raindrops of particles in the magnetosphere},
  author={Chen, X and Yan, Y and Han, JL and Wang, C and Wang, PF and Jing, WC and Lee, KJ and Zhang, B and Xu, RX and Wang, T and others},
  journal={Nature Astronomy},
  volume={7},
  number={10},
  pages={1235--1244},
  year={2023},
  publisher={Nature Publishing Group UK London}
}

@article{yan2024dwarf,
  title={Dwarf pulses of 10 pulsars detected by FAST},
  author={Yan, Yi and Han, JL and Zhou, DJ and Xie, L and Kou, FF and Wang, PF and Wang, C and Wang, T},
  journal={The Astrophysical Journal},
  volume={965},
  number={1},
  pages={25},
  year={2024},
  publisher={IOP Publishing}
}

@article{kramer2006periodically,
  title={A periodically active pulsar giving insight into magnetospheric physics},
  author={Kramer, Michael and Lyne, Andrew G and O'Brien, JT and Jordan, Christine A and Lorimer, Duncan R},
  journal={Science},
  volume={312},
  number={5773},
  pages={549--551},
  year={2006},
  publisher={American Association for the Advancement of Science}
}

@article{young2014apparent,
  title={On the apparent nulls and extreme variability of PSR J1107- 5907},
  author={Young, NJ and Weltevrede, P and Stappers, BW and Lyne, AG and Kramer, M},
  journal={Monthly Notices of the Royal Astronomical Society},
  volume={442},
  number={3},
  pages={2519--2533},
  year={2014},
  publisher={Oxford University Press}
}

@article{johnston2024thousand,
  title={The Thousand-Pulsar-Array programme on MeerKAT--XIV. On the high linearly polarized pulsar signals},
  author={Johnston, Simon and Mitra, Dipanjan and Keith, Michael J and Oswald, Lucy S and Karastergiou, Aris},
  journal={Monthly Notices of the Royal Astronomical Society},
  volume={530},
  number={4},
  pages={4839--4849},
  year={2024},
  publisher={Oxford University Press}
}

@article{basu2016meterwavelength,
  title={Meterwavelength single-pulse polarimetric emission survey. II. The phenomenon of drifting subpulses},
  author={Basu, Rahul and Mitra, Dipanjan and Melikidze, George I and Maciesiak, Krzysztof and Skrzypczak, Anna and Szary, Andrzej},
  journal={The Astrophysical Journal},
  volume={833},
  number={1},
  pages={29},
  year={2016},
  publisher={IOP Publishing}
}

@article{basu2020periodic,
  title={Periodic modulation: newly emergent emission behavior in pulsars},
  author={Basu, Rahul and Mitra, Dipanjan and Melikidze, Giorgi I},
  journal={The Astrophysical Journal},
  volume={889},
  number={2},
  pages={133},
  year={2020},
  publisher={IOP Publishing}
}

@article{han2021fast,
  title={The FAST Galactic Plane Pulsar Snapshot survey: I. Project design and pulsar discoveries⋆},
  author={Han, JL and Wang, Chen and Wang, PF and Wang, Tao and Zhou, DJ and Sun, Jing-Hai and Yan, Yi and Su, Wei-Qi and Jing, Wei-Cong and Chen, Xue and others},
  journal={Research in Astronomy and Astrophysics},
  volume={21},
  number={5},
  pages={107},
  year={2021},
  publisher={IOP Publishing}
}

@article{van2002null,
  title={Null-induced mode changes in PSR B0809+ 74},
  author={Van Leeuwen, AGJ and Kouwenhoven, MLA and Ramachandran, R and Rankin, JM and Stappers, BW},
  journal={Astronomy \& Astrophysics},
  volume={387},
  number={1},
  pages={169--178},
  year={2002},
  publisher={EDP Sciences}
}

@article{janssen2004intermittent,
  title={Intermittent nulls in PSR B0818-13, and the subpulse-drift alias mode},
  author={Janssen, GH and van Leeuwen, Joeri},
  journal={Astronomy \& Astrophysics},
  volume={425},
  number={1},
  pages={255--261},
  year={2004},
  publisher={EDP Sciences}
}

@article{redman2005pulsar,
  title={Pulsar PSR B2303+ 30: a single system of drifting subpulses, moding and nulling},
  author={Redman, Stephen L and Wright, Geoffrey AE and Rankin, Joanna M},
  journal={Monthly Notices of the Royal Astronomical Society},
  volume={357},
  number={3},
  pages={859--872},
  year={2005},
  publisher={The Royal Astronomical Society}
}

@article{backer1970pulsar,
  title={Pulsar nulling phenomena},
  author={Backer, DC},
  journal={Nature},
  volume={228},
  number={5266},
  pages={42--43},
  year={1970},
  publisher={Nature Publishing Group UK London}
}

@article{rankin1986toward,
  title={Toward an empirical theory of pulsar emission. III-Mode changing, drifting subpulses, and pulse nulling},
  author={Rankin, Joanna M},
  journal={Astrophysical Journal, Part 1 (ISSN 0004-637X), vol. 301, Feb. 15, 1986, p. 901-922. Research supported by the Research Corporation and University of Vermont.},
  volume={301},
  pages={901--922},
  year={1986}
}

@article{wang2007pulsar,
  title={Pulsar nulling and mode changing},
  author={Wang, Nina and Manchester, RN and Johnston, S},
  journal={Monthly Notices of the Royal Astronomical Society},
  volume={377},
  number={3},
  pages={1383--1392},
  year={2007},
  publisher={Blackwell Publishing Ltd Oxford, UK}
}

@article{ng2020discovery,
  title={The Discovery of Nulling and Mode-switching Pulsars with CHIME/Pulsar},
  author={Ng, C and Wu, B and Ma, M and Ransom, SM and Naidu, A and Fonseca, E and Boyle, PJ and Brar, C and Cubranic, D and Demorest, PB and others},
  journal={The Astrophysical Journal},
  volume={903},
  number={2},
  pages={81},
  year={2020},
  publisher={IOP Publishing}
}

@article{herfindal2007periodic,
  title={Periodic nulls in the pulsar B1133+ 16},
  author={Herfindal, Jeffrey L and Rankin, Joanna M},
  journal={Monthly Notices of the Royal Astronomical Society},
  volume={380},
  number={2},
  pages={430--436},
  year={2007},
  publisher={The Royal Astronomical Society}
}

@article{wang2020two,
  title={The two emission states of PSR B1534+ 12},
  author={Wang, SQ and Hobbs, G and Wang, JB and Manchester, R and Wang, N and Zhang, SB and Feng, Y and Wang, W-Y and Li, D and Dai, S and others},
  journal={The Astrophysical Journal Letters},
  volume={902},
  number={1},
  pages={L13},
  year={2020},
  publisher={IOP Publishing}
}

@article{helfand1975observations,
  title={Observations of pulsar radio emission. III-Stability of integrated profiles},
  author={Helfand, DJ and Manchester, RN and Taylor, JH},
  journal={Astrophysical Journal, vol. 198, June 15, 1975, pt. 1, p. 661-670.},
  volume={198},
  pages={661--670},
  year={1975}
}

@article{rathnasree1995approach,
  title={On the approach to stability of pulsar average profiles},
  author={Rathnasree, N and Rankin, Joanna M},
  journal={Astrophysical Journal},
  volume={452},
  pages={814--818},
  year={1995},
  publisher={The University of Chicago Press for the American Astronomical Society}
}

@article{liu2012profile,
  title={Profile-shape stability and phase-jitter analyses of millisecond pulsars},
  author={Liu, K and Keane, EF and Lee, KJ and Kramer, M and Cordes, JM and Purver, MB},
  journal={Monthly Notices of the Royal Astronomical Society},
  volume={420},
  number={1},
  pages={361--368},
  year={2012},
  publisher={The Royal Astronomical Society}
}

@article{backer1970peculiar,
  title={Peculiar pulse burst in PSR 1237+ 25},
  author={Backer, DC},
  journal={Nature},
  volume={228},
  number={5278},
  pages={1297--1298},
  year={1970},
  publisher={Nature Publishing Group UK London}
}

@article{rankin1974individual,
  title={Individual pulse polarization properties of three pulsars},
  author={Rankin, John M and Campbell, DB and Backer, DC},
  journal={Astrophysical Journal, Vol. 188, pp. 609-614 (1974)},
  volume={188},
  pages={609--614},
  year={1974}
}

@article{rankin2013drifting,
  title={Drifting, moding and nulling: another look at pulsar B1918+ 19},
  author={Rankin, Joanna M and Wright, Geoffrey AE and Brown, Andrew M},
  journal={Monthly Notices of the Royal Astronomical Society},
  volume={433},
  number={1},
  pages={445--456},
  year={2013},
  publisher={The Royal Astronomical Society}
}

@article{lorimer2006parkes,
  title={The Parkes Multibeam Pulsar Survey--VI. Discovery and timing of 142 pulsars and a Galactic population analysis},
  author={Lorimer, Duncan R and Faulkner, AJ and Lyne, AG and Manchester, Richard N and Kramer, M and McLaughlin, MA and Hobbs, G and Possenti, A and Stairs, IH and Camilo, F and others},
  journal={Monthly Notices of the Royal Astronomical Society},
  volume={372},
  number={2},
  pages={777--800},
  year={2006},
  publisher={Blackwell Publishing Ltd Oxford, UK}
}

@article{yang2014new,
  title={A new method to analyse pulsar nulling phenomenon},
  author={Yang, AiYuan and Han, JinLin and Wang, Na},
  journal={Science China Physics, Mechanics \& Astronomy},
  volume={57},
  number={8},
  pages={1600--1606},
  year={2014},
  publisher={Springer}
}

@article{di2018considerations,
  title={Considerations for a Multi-beam Multi-purpose Survey with FAST},
  author={Di Li, Pei Wang and Qian, Lei and Krco, Marko and Jiang, Peng and Yue, Youling and Jin, Chenjin and Zhu, Yan and Pan, Zhichen and Nan, Rendong},
  journal={IEEE Microwave},
  volume={19},
  pages={112--119},
  year={2018}
}

@article{van2011dspsr,
  title={DSPSR: digital signal processing software for pulsar astronomy},
  author={van Straten, Wꎬ and Bailes, M},
  journal={Publications of the Astronomical Society of Australia},
  volume={28},
  number={1},
  pages={1--14},
  year={2011},
  publisher={Cambridge University Press}
}

@article{hotan2004psrchive,
  title={PSRCHIVE and PSRFITS: an open approach to radio pulsar data storage and analysis},
  author={Hotan, Aidan W and van Straten, Willem and Manchester, RN},
  journal={Publications of the Astronomical Society of Australia},
  volume={21},
  number={3},
  pages={302--309},
  year={2004},
  publisher={Cambridge University Press}
}

@article{zhao2017tmrt,
  title={TMRT observations of 26 pulsars at 8.6 GHz},
  author={Zhao, Ru-Shuang and Wu, Xin-Ji and Yan, Zhen and Shen, Zhi-Qiang and Manchester, RN and Qiao, Guo-Jun and Xu, Ren-Xin and Wu, Ya-Jun and Zhao, Rong-Bing and Li, Bin and others},
  journal={The Astrophysical Journal},
  volume={845},
  number={2},
  pages={156},
  year={2017},
  publisher={IOP Publishing}
}

@article{bhattacharyya2008results,
  title={Results from multifrequency observations of PSR B0826- 34},
  author={Bhattacharyya, B and Gupta, Y and Gil, J},
  journal={Monthly Notices of the Royal Astronomical Society},
  volume={383},
  number={4},
  pages={1538--1550},
  year={2008},
  publisher={Blackwell Publishing Ltd Oxford, UK}
}

@article{young2015long,
  title={Long-term observations of three nulling pulsars},
  author={Young, NJ and Weltevrede, Patrick and Stappers, BW and Lyne, AG and Kramer, Michael},
  journal={Monthly Notices of the Royal Astronomical Society},
  volume={449},
  number={2},
  pages={1495--1504},
  year={2015},
  publisher={Oxford University Press}
}

@article{rankin1983toward,
  title={Toward an empirical theory of pulsar emission. I Morphological taxonomy.},
  author={Rankin, Joanna M},
  journal={Astrophysical Journal, Part 1 (ISSN 0004-637X), vol. 274, Nov. 1, 1983, p. 333-368. Research supported by the University of Vermont.},
  volume={274},
  pages={333--368},
  year={1983}
}

@article{rankin1993toward,
  title={Toward an empirical theory of pulsar emission. VI-The geometry of the conal emission region},
  author={Rankin, Joanna M},
  journal={Astrophysical Journal},
  volume={405},
  pages={285--297},
  year={1993},
  publisher={The University of Chicago Press for the American Astronomical Society}
}

@article{jones2011instabilities,
  title={Instabilities, nulls and subpulse drift in radio pulsars},
  author={Jones, PB},
  journal={Monthly Notices of the Royal Astronomical Society},
  volume={414},
  number={1},
  pages={759--769},
  year={2011},
  publisher={Blackwell Publishing Ltd Oxford, UK}
}

@article{biggs1992analysis,
  title={An analysis of radio pulsar nulling statistics},
  author={Biggs, James D},
  journal={Astrophysical Journal, Part 1 (ISSN 0004-637X), vol. 394, no. 2, Aug. 1, 1992, p. 574-580.},
  volume={394},
  pages={574--580},
  year={1992}
}

@article{gajjar2014frequency,
  title={Frequency independent quenching of pulsed emission},
  author={Gajjar, Vishal and Joshi, Bhal Chandra and Kramer, Michael and Karuppusamy, Ramesh and Smits, Roy},
  journal={The Astrophysical Journal},
  volume={797},
  number={1},
  pages={18},
  year={2014},
  publisher={IOP Publishing}
}




\appendix


\bsp	
\label{lastpage}
\end{document}